\documentclass{article}

\PassOptionsToPackage{numbers,compress}{natbib}
\PassOptionsToPackage{table,xcdraw}{xcolor}

\usepackage[preprint]{neurips_2026}

\usepackage[utf8]{inputenc} 
\usepackage[T1]{fontenc}    
\usepackage[hidelinks]{hyperref} 
\usepackage{url}            
\usepackage{booktabs}       
\usepackage{amsfonts}       
\usepackage{nicefrac}       
\usepackage{microtype}      
\usepackage[table,xcdraw]{xcolor} 
\usepackage{graphicx}       
\usepackage{subcaption}     
\usepackage{enumitem}       
\usepackage{multirow}       
\usepackage{colortbl}       
\usepackage{siunitx}        
\usepackage{pifont}         
\usepackage{amsmath}        
\usepackage{amssymb}        
\usepackage{mathtools}      
\usepackage{amsthm}         
\usepackage{makecell}       
\usepackage{fontawesome5}   
\usepackage{xspace}         
\usepackage{xurl}           
\usepackage[capitalize,noabbrev]{cleveref} 
\usepackage{arydshln}

\newcommand{\cmark}{\textcolor{green!60!black}{\ding{51}}}
\newcommand{\xmark}{\textcolor{red}{\ding{55}}}

\newcommand{\sysname}{\textsc{PostEDA-Bench}\xspace}

\definecolor{gold}{RGB}{255,215,0}
\definecolor{silver}{RGB}{192,192,192}
\definecolor{bronze}{RGB}{205,127,50}
\definecolor{orange}{RGB}{255,220,147}
\definecolor{pink}{RGB}{255,215,201}

\theoremstyle{plain}

\theoremstyle{definition}

\theoremstyle{remark}

\title{Bridging the Last Mile of Circuit Design: \textsc{\textcolor[HTML]{FF9855}{Post}\textcolor[HTML]{85C5F2}{E}\textcolor[HTML]{9DCE81}{D}\textcolor[HTML]{FBCF64}{A}-Bench}, a Hierarchical Benchmark for PPA Convergence and DRC Fixing}

\author{%
  Pengju Liu$^{1}$, Nuo Xu$^{1}$, Jinwei Tang$^{1}$, Yu Cao$^{1}$, Caiwen Ding$^{1}$ \\
  $^{1}$University of Minnesota \\
  \texttt{liu03486@umn.edu}
}

\begin{document}

\maketitle

\begin{abstract}
LLM-based agents are increasingly applied to the ``last mile'' of Electronic Design Automation (EDA): repairing residual sign-off Design Rule Check (DRC) violations and converging Power--Performance--Area (PPA) targets after tool runs. Existing EDA-LLM benchmarks, however, omit DRC fixing entirely and rely on flat hierarchies tied to a single toolchain. We introduce \textsc{PostEDA-Bench}, a hierarchical benchmark with $145$ tasks across \textbf{DRC-Essential}, \textbf{DRC-Reasoning}, \textbf{PPA-Mono}, and \textbf{PPA-Multi}, supported by EDA toolchains with machine-checkable evaluation. Across nine commercial and open-source LLMs under multiple agent scaffolds, we find that agents handle synthetic DRC-Essential and single-objective PPA-Mono reasonably well but degrade sharply on the more practical DRC-Reasoning (best SR $54.67\%$) and PPA-Multi (best SR $20.00\%$); vision augmentation improves pooled DRC success rates; and trade-off reasoning---rather than knob knowledge---is the dominant PPA-Multi bottleneck. Our benchmark is open-sourced at \url{https://github.com/pengjas/posteda-bench}.
\end{abstract}

\section{Introduction}
Large language model (LLM)--based agentic systems have demonstrated strong performance in code generation\cite{hong2024metagpt, zhang-etal-2024-codeagent, chen2024teaching}, robotics and embodied AI\cite{yao2023react, palm}, and mathematical reasoning\cite{toolformer, pal}. These advances have motivated recent work on LLM-driven agents for automating power--performance--area (PPA) convergence in electronic design automation (EDA) flows \cite{wu2024chateda, wu2025EDAid, wang2025mcp4edallmpoweredmodelcontext, lu2025autoedaenablingedaflow, jiang2024iicpilotintelligentintegratedcircuit, Fu2023GPT4AIGChip, liu2024chipnemodomainadaptedllmschip, ghose2025orfsagenttoolusingagentschip, lu2025autoedaenablingedaflow}, where engineers often spend significant time iterating after tool runs to meet PPA targets.

Despite this progress, existing EDA-agent benchmarks still \textbf{fall short of what is needed to evaluate practical post-EDA closure}. As summarized in \cref{tab:benchmark-comparison}, most benchmarks either focus on \textbf{NLP-to-script generation rather than target-driven PPA optimization}, contain only a small number of optimization tasks, use flat or coarse task hierarchies, or are tied to a single toolchain, limiting both diagnostic resolution and generality. More importantly, \textbf{none of them evaluates design rule check (DRC) fixing}, although residual sign-off DRC violations remain a time-critical ``last-mile'' workload even after modern routing. These gaps make a unified and realistic benchmark for post-EDA PPA optimization and DRC repair both \textbf{urgent and necessary}.
\begin{table*}[]
\centering
\caption{\textbf{Comparison with related benchmarks.}}
\label{tab:benchmark-comparison}
\resizebox{\textwidth}{!}{%
\begin{tabular}{lcccccccc}
\Xhline{1.2pt}
\rowcolor[HTML]{EFEFEF} 
Benchmark       & \#Tasks & \begin{tabular}[c]{@{}c@{}}Open\\-sourced\end{tabular} &Objective & Hierarchy  & EDA Toolchains  & DRC Fix \\ \hline
ChatEDA-bench \cite{wu2024chateda}   & 50            & \cmark    &   NLP$\rightarrow$scripts  &     Simple--Complex                        &    OpenROAD             &     \xmark                \\
GPT4AIGChip \cite{Fu2023GPT4AIGChip}     & 6            & \xmark     &  Codegen\&PPA Opt.   &    Flat                         &    Vivado HLS tools             &   \xmark                   \\
Chipnemo \cite{liu2024chipnemodomainadaptedllmschip}        &      959        &   \xmark                 &NLP$\rightarrow$scripts  & Easy--Medium--Hard                 &     n/a            &          \xmark            \\
IICPilot \cite{jiang2024iicpilotintelligentintegratedcircuit}        & 4            & \xmark                   & PPA Opt.&   Flat               &       OpenROAD         &      \xmark                \\
iEDA-bench \cite{wu2025EDAid}      & 50            & \xmark    & NLP$\rightarrow$scripts  &   Simple--Complex                  &  OpenROAD         &                        \xmark              \\
MCP4EDA \cite{wang2025mcp4edallmpoweredmodelcontext}         & 10            & \cmark    &  NLP$\rightarrow$scripts  &    Flat               &  OpenROAD        &                 \xmark                      \\
ORFS-agent \cite{ghose2025orfsagenttoolusingagentschip}      &      6        &   \cmark   & PPA Opt. &   Flat                            &   OpenROAD              &        \xmark              \\
AutoEDA \cite{lu2025autoedaenablingedaflow}         & 100            &  \xmark  & NLP$\rightarrow$scripts & Different--stage in EDA                                &    DC\&Innovus             & \xmark                     \\ \hline
\rowcolor[HTML]{EFEFEF}
\textbf{\textsc{PostEDA-Bench}} &      145        & \cmark   &  PPA Opt. & \begin{tabular}[c]{@{}c@{}}Multi-dimension\&\\Multi-level\end{tabular}                &   \begin{tabular}[c]{@{}c@{}}OpenROAD\&\\DC\&Innovus\end{tabular}        &                   \cmark                   \\ \Xhline{1.2pt}
\multicolumn{7}{l}{$\dagger$: NLP$\rightarrow$scripts: generates EDA scripts from natural language; it runs tools but doesn’t optimize PPA. } \\
\multicolumn{7}{l}{$\S$: PPA Opt. denotes PPA Optimization. $\star$: OpenROAD is open-source, whereas DC and Innovus are commercial tools.} \\
\multicolumn{7}{l}{$\ddagger$: DC and Innovus are only used to generate part of DRC-Bench tasks in \textsc{PostEDA-Bench}.}
\end{tabular}%
}
\vspace{-10pt}
\end{table*}

To fill this gap, we introduce \textbf{\textsc{PostEDA-Bench}}, a hierarchical benchmark for evaluating LLM-based agents on post-EDA closure: \textbf{\textsc{PostEDA-Bench}:} 145 tasks across two dimensions, \textbf{DRC-Bench} and \textbf{PPA-Bench}, evaluating whether an agent can (i) repair sign-off DRC violations and (ii) drive designs to specified PPA targets, supported by both open-source (\textsc{OpenROAD}) and commercial (\textsc{DC}+\textsc{Innovus}) toolchains with machine-checkable evaluation. \textbf{DRC-Bench:} splits into DRC-Essential, three levels probing rule knowledge, contextual robustness, and cascaded multi-violation fixing on synthetic cases, and DRC-Reasoning, three levels of practical residual post-flow violations that demand geometric reasoning, stratified by minimum required edit steps. \textbf{PPA-Bench:} splits into PPA-Mono (Performance, Power, Area), with single- and multi-knob perturbation levels plus a Performance re-architecture level, and PPA-Multi, with 2D and 3D Pareto-targeted tasks that test trade-off reasoning under constraint floors.

Using \textsc{PostEDA-Bench}, we evaluate commercial and open-source models, frontier and small backbones, and text-only versus vision--language variants under several agent scaffolds (ReAct, Proposer--Critic, ORFS-Agent, Reflexion and ToT). \textbf{Our key findings are as follows:}
\begin{itemize}[leftmargin=*, label={$\triangleright$}, nosep]
    \item \textbf{Synthetic vs.\ practical asymmetry.} Agents perform reasonably well on synthetic DRC-Essential (best SR $95.00\%$) and single-objective PPA-Mono (best SR $64.56\%$), but degrade sharply on the more practical DRC-Reasoning (best SR $54.67\%$) and PPA-Multi (best SR $20.00\%$), where post-flow geometric reasoning and multi-objective trade-offs dominate.
    \item \textbf{Vision augmentation enhances DRC-Bench.} Adding a layout-image channel maintains or improves SR/VRR, with positive pooled SR effects on DRC-Essential and DRC-Reasoning. Individual gains are not uniformly significant; paired bootstrap CIs appear in \cref{app:variance}.
    \item \textbf{Trade-off reasoning is the PPA-Multi bottleneck.} Several backbones produce \emph{negative} NIS on PPA-Multi by regressing constraint metrics while chasing the violated objective, indicating they greedily optimize one PPA dimension rather than balancing competing targets.
\end{itemize}

\section{Related Work}
\noindent\textbf{DRC fixing.}
Design rule checking verifies layout manufacturability, yet residual sign-off errors and cascaded fixes often remain after automated repair. This motivates evaluating agents on rule understanding, layout inspection, geometric editing, and iterative DRC checking; additional background is in \cref{app:related-work}. \noindent\textbf{PPA optimization.}
Power, performance, and area are coupled objectives, and automated flows often miss tight targets despite many configuration, constraint, and design-level knobs. PPA-Bench targets this last-mile loop by testing whether agents can use tool feedback for single-objective improvement and multi-objective trade-offs; extended context is in \cref{app:related-work}.

\section{\textsc{PostEDA-Bench}}

\begin{figure}[t]
    \centering
    \begin{minipage}[c]{0.61\linewidth}
        \centering
        \captionof{table}{\textbf{Statistics of \textsc{PostEDA-Bench}}}
        \label{tab:benchmark-statistics}
        \resizebox{\linewidth}{!}{%
        \renewcommand{\arraystretch}{1.1}
        \begin{tabular}{lc}
        \hline
        \rowcolor[HTML]{C0C0C0}
        \textbf{Statistic} & \textbf{Number} \\ \hline \hline

        \rowcolor[HTML]{EFEFEF}
        Tasks & 145 \\

        \rowcolor[HTML]{EFEFEF}
        Std. cells per source design & min 301; avg 16,505; max 95,061 \\

        \rowcolor[HTML]{EFEFEF}
        Covered layers in DRC-Bench & N(P)Well -- M9 \\

        \rowcolor[HTML]{EFEFEF}
        DRC-Essential rules (before pruning) & 168 \\

        \rowcolor[HTML]{EFEFEF}
        Human fix steps (DRC-Reasoning) & min 1; max 8 \\

        \rowcolor[HTML]{EFEFEF}
        DRC runtime$^\dagger$ & min 10\,s; max 3\,h \\

        \rowcolor[HTML]{EFEFEF}
        Needed Tunable parameters$^{\S}$ & min 1; max 7 \\

        \rowcolor[HTML]{EFEFEF}
        Mono-Perf tasks requiring re-architecture$^\star$ & 4 \\

        \hline
        \end{tabular}%
        }
        \vspace{2pt}
        \parbox{\linewidth}{\footnotesize
        $\dagger$: DRC runtime is the wall-clock time for one signoff DRC check; it reflects design-scale diversity and is not agent or human repair time.\\
        $\S$: Needed tunable parameters are minimum critical knobs; agents may edit all OpenROAD parameters.\\
        $\star$: Mono-Perf re-architecture tasks require RTL/micro-architectural changes beyond parameter tuning.}
    \end{minipage}\hfill
    \begin{minipage}[c]{0.39\linewidth}
        \centering
        \includegraphics[width=\linewidth]{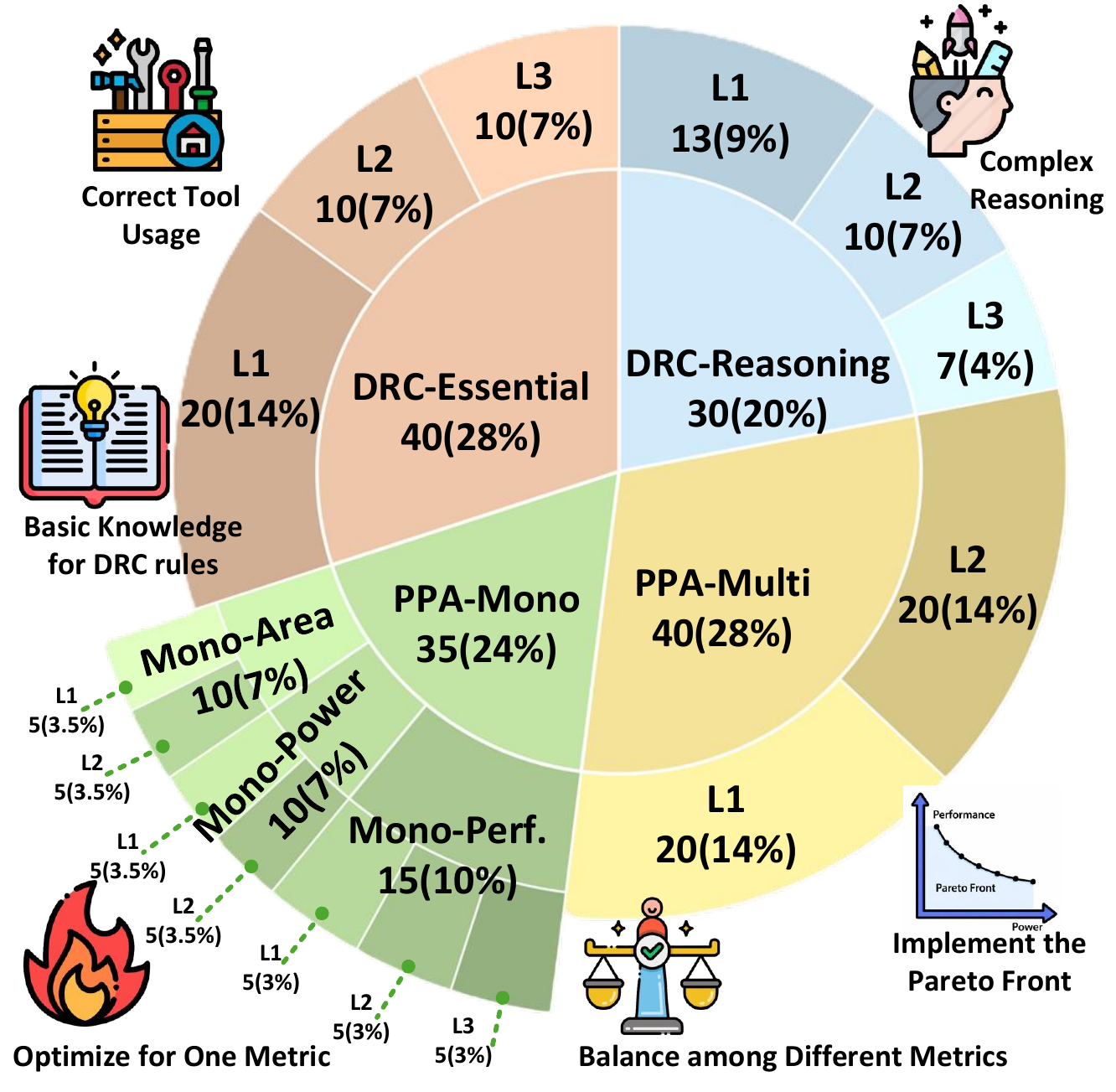}
        \captionof{figure}{\textbf{Overview of \sysname composition.}}
        \label{fig:benchmark}
    \end{minipage}
    \vspace{-10pt}
\end{figure}

\textbf{Preliminary Source Design Collection.}
As shown on the left of \cref{fig:workflow}, \sysname starts from three public RTL sources: \textsc{RTLLM} v2~\cite{lu2024rtllm} (50 designs), \textsc{VerilogEval}-Human~\cite{liu2023verilogeval} (156 problems), and \textsc{OpenCores}~\cite{opencores} (1,283 listed IP projects), covering arithmetic circuits, controllers, and community IP blocks. 
We map retained designs to ASAP7~\cite{clark2016asap7} and keep only candidates satisfying four auditable criteria: (i) \textbf{parse and synthesis validity}, i.e., a well-defined top module, valid syntax, and successful gate-level synthesis; (ii) \textbf{non-trivial scale}, with more than 300 instantiated standard cells; (iii) \textbf{cross-source deduplication} by top-module names, port signatures, and post-synthesis structural similarity, merging designs with more than 90\% gate-level overlap; and (iv) \textbf{release and audit eligibility}, requiring documented provenance, license, selected top module, and derived artifacts. 
Aggregate statistics are reported in \cref{tab:benchmark-statistics}; \cref{app:source-design-protocol} details deduplication, contamination control, and the release package.

\textbf{Contamination control.} We mitigate contamination through task-specific geometric edits and PPA targets verified by fresh tool runs; task design and verification are detailed in \cref{app:contamination}.
\textbf{Reproducibility.} Most tasks are constructed manually by the authors, with scripts assisting artifact generation and parameter sweeps. Both suites provide fixed initial artifacts, prompts, metadata, pinned tool setup, and evaluation drivers so that evaluation can be repeated without recreating the authors' manual decisions. All released tasks can be scored with open-source tools. Construction provenance, release contents, and versioning are documented in \cref{app:source-design-protocol,app:reproducibility,app:ppa-reproducibility}.

\subsection{DRC-Bench}
\label{sec:drc-bench}
DRC-Bench evaluates last-mile sign-off DRC repair, a workload absent from existing EDA-LLM benchmarks (\cref{tab:benchmark-comparison}), through two branches: \textbf{DRC-Essential}, which isolates rule knowledge, layout-tool use, context robustness, and cascaded fixing on synthetic cases, and \textbf{DRC-Reasoning}, which targets residual post-flow violations requiring geometric reasoning.


\subsubsection{Task Specifications}



\begin{figure}[bp]
  \centering

  \begin{subfigure}[b]{0.49\columnwidth}
    \centering
    \includegraphics[width=\linewidth]{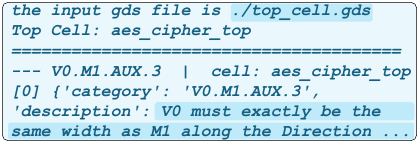}
    \caption{Sample prompt in DRC-Bench.}
    \label{fig:left}
  \end{subfigure}
  \hfill
  \begin{subfigure}[b]{0.49\columnwidth}
    \centering
    \includegraphics[width=\linewidth]{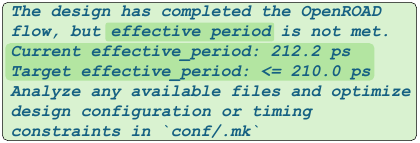}
    \caption{Sample prompt in PPA-Bench.}
    \label{fig:right}
  \end{subfigure}

  \caption{\textbf{Example prompts used to drive the agent in DRC-Bench and PPA-Bench.}}
  \label{fig:two-subfig-horizontal}
\end{figure}

\textbf{Input.} Each task provides two artifacts: the target \texttt{.gds} layout and a natural-language prompt (\cref{fig:left}). The prompt includes the initial sign-off DRC violations, with each entry specifying the rule identifier (e.g., \texttt{V0.M1.AUX.3}), a short rule description, the geometry type (\texttt{edge-pair}, \texttt{polygon}, or \texttt{edge}), and layout coordinates. \textbf{Metrics.} A run is \textbf{successful} when the final layout is DRC-clean under sign-off DRC, and we report the \textbf{Success Rate (SR)}, computed as the mean of the success indicator over $5$ independent runs per task. To capture partial progress, we additionally report the \textbf{Violation Reduction Rate (VRR)}:
\begin{equation}
    \mathrm{VRR} = \frac{\max\!\left(0,\, E_{\text{initial}} - E_{\text{final}}\right)}{E_{\text{initial}}}\times 100\%,
\end{equation}
where $E_{\text{initial}}$ and $E_{\text{final}}$ denote the total violation count before and after the agent's edits. The choice of sign-off tool and DRC deck is documented in \cref{app:reproducibility}. A post-hoc LVS audit is in \cref{app:drc-functional-audit}. \textbf{Tools.} Agents can query layout context textually or visually, edit GDS geometry and placement, and rerun sign-off DRC through a compact layout-viewer-style tool set; full schemas are in \cref{app:tools}.

\subsubsection{DRC-Essential}
\label{sec:drc-essential}
\begin{figure*}[tbp]  
    \centering  
    \includegraphics[width=\linewidth]{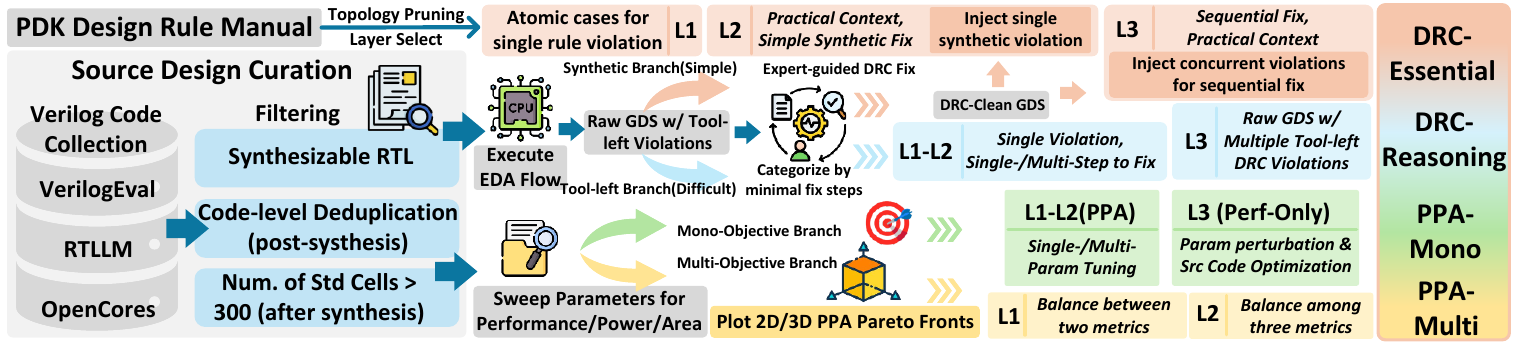}  
    \caption{
\textbf{Overview of \sysname construction process.}
}
    \label{fig:workflow}  
     \vspace{-10pt}
\end{figure*}

DRC-Essential has three levels (light-apricot block in \cref{fig:workflow}), each isolating a primitive competency. \textbf{L1 (Atomic rule understanding).} L1 uses atomic single-rule violations with only the required shapes. Cases are hand-drawn from the ASAP7 design-rule manual or generated by per-task scripts (\texttt{create\_errors.py}; \cref{app:rule-coverage}). We group \emph{topologically equivalent} rules---those differing only in layer assignment---into one representative per class, balanced to one case per layer. \textbf{L2 (Contextual comprehensiveness).} L2 embeds one L1-style violation in realistic layout context, forcing agents to filter relevant geometry from clutter. Construction: \emph{(i)} run a full EDA flow---\textsc{OpenROAD-flow-scripts} or \textsc{Synopsys DC} +\textsc{Cadence Innovus}---on a source-pool design; \emph{(ii)} have an engineer repair residuals to a DRC-clean reference; \emph{(iii)} inject one L1-style violation whose fix touches only the offending shapes. L2 thus isolates context filtering without adding geometric reasoning. \textbf{L3 (Sequential fixing).} L3 tests whether agents can resolve interacting violation chains. For each DRC-clean design we inject $5$--$15$ L1-style violations across rule families, mainly width/via (e.g., \texttt{M2.W.1}, \texttt{M3.W.1}, etc.). Sites are co-located so edits interact; each violation is single-step in isolation, but agents must order edits and re-query DRC since fixes can trigger or eliminate nearby ones.




\subsubsection{DRC-Reasoning}
\label{sec:drc-reasoning}
\textbf{DRC-Reasoning targets practical residual violations that remain after full EDA flows and built-in repair} (blue block in \cref{fig:workflow}; \cref{app:innovus-repair-study} quantifies the residual gap for Innovus). These cases are harder than DRC-Essential because they require non-trivial geometric reasoning. We partition them into three levels by the \emph{minimum number of agent editing steps} required (one step = one \texttt{add\_shape}/\texttt{change\_shape}/\texttt{move\_cell} call; inspection not counted). Labels come from an experienced physical-design engineer following the protocol in \cref{app:step-protocol}, with a delayed self-consistency pass. Two physical-design engineers completed cross-verification of the labels under this protocol. \textbf{L1--L2 (Single- vs. multi-step fixes).} We sweep \textsc{OpenROAD} and \textsc{DC}+\textsc{Innovus} for post-flow residual violations across open-source and industrial settings. A violation is \textbf{L1} if its repair takes one editing step, \textbf{L2} if multiple (no upper cap). When unrelated violations co-exist, we isolate the target by cropping or temporarily fixing surroundings, so L1/L2 cleanly probe single-violation reasoning. \textbf{L3 (Multi-violation debugging).} L3 reproduces the practical multi-violation scenario: each task is a raw post-flow GDS with several real violations that the agent must iteratively debug and resolve. L3 designs may overlap with those seeding L1/L2 (different post-flow configs of the same source), but the per-task violation set is disjoint by construction.



\subsection{PPA-Bench}
Beyond DRC, post-tool closure often fails because designs miss PPA (power, performance, and area) constraints.
PPA-Bench contains two subsets: \textbf{PPA-Mono} optimizes one metric (performance, power, or area), while \textbf{PPA-Multi} evaluates trade-offs among multiple metrics.

\subsubsection{Task Specifications}
\textbf{Inputs.} Each task is a self-contained OpenROAD project with Verilog sources (\texttt{src/}), configs (\texttt{config.mk}, \texttt{constraints.sdc}), and previous-run artifacts (\texttt{logs/}, \texttt{reports/}, \texttt{objects/}, \texttt{results/}). The prompt summarizes current PPA, target thresholds, constraint floors, and the \texttt{FlowVariables.md} knob list (\cref{fig:right}); ground-truth metadata is access-controlled in \cref{app:ppa-manifest}. \textbf{Metrics.} A run is \textbf{successful} when every target metric is met and the modified RTL passes the task's applicable \textbf{testbench-based functional checks} (\cref{app:func-equiv}); \textbf{Success Rate (SR)} is the mean success fraction over $5$ temperature-$0$ runs per task. We also report the \textbf{Normalized Improvement Score (NIS)} for partial progress. For each constrained metric $i \in \mathcal{M}$, let $\mathcal{A} = \{i : M_i^{\mathrm{init}} > M_i^{\mathrm{tgt}}\}$ be the violated metrics to improve and $\mathcal{B} = \{i : M_i^{\mathrm{init}} \le M_i^{\mathrm{tgt}}\}$ be constraints that must not regress; the per-metric score is
\begin{equation}
\footnotesize
\label{eq:ppa-si}
s_i =
\begin{cases}
\min\!\big(1,\, r_i\big), & i \in \mathcal{A},\\[2pt]
1, & i \in \mathcal{B} \text{ and } M_i^{\mathrm{fin}} \le M_i^{\mathrm{tgt}},\\[2pt]
1 - r_i, & i \in \mathcal{B} \text{ and } M_i^{\mathrm{fin}} > M_i^{\mathrm{tgt}},
\end{cases}
\quad \text{where } r_i = \frac{M_i^{\mathrm{fin}} - M_i^{\mathrm{init}}}{M_i^{\mathrm{tgt}} - M_i^{\mathrm{init}}}.
\end{equation}
The aggregate score is then
\begin{equation}
\footnotesize
\label{eq:score-ppa}
\mathrm{NIS} =
\begin{cases}
0, & \mathcal{A} \neq \emptyset \text{ and } M_i^{\mathrm{fin}} \ge M_i^{\mathrm{init}}\ \forall\, i \in \mathcal{A},\\[2pt]
\dfrac{1}{|\mathcal{M}|}\displaystyle\sum_{i \in \mathcal{M}} s_i, & \text{otherwise.}
\end{cases}
\end{equation}
All metrics are lower-is-better: $r_i = 1$ means the target is reached, $r_i \in (0,1)$ means partial progress, and $r_i < 0$ means regression. For single-objective tasks ($|\mathcal{M}|=1$) we clip $s_1$ to $[0,1]$; for multi-objective tasks, negative scores penalize regression, and Eq.\,\ref{eq:score-ppa}'s zero-floor rule prevents trade-off exploits that improve only constraint metrics. Per-task scores are averaged within each class. \textbf{Run protocol.} Each run is bounded by a tool-call iteration cap, treated as an evaluation hyperparameter and varied in the iteration ablation. \textbf{Tools.} Agents inspect files, edit configuration/constraints/RTL, run OpenROAD in an isolated container, and parse PPA reports through four sandboxed tools; full schemas and locked variables are in \cref{app:ppa-tools}.

\subsubsection{PPA-Mono}
PPA-Mono evaluates single-objective optimization across \textbf{Mono-Performance}, \textbf{Mono-Power}, and \textbf{Mono-Area} (green block in \cref{fig:workflow}), with no constraints on the other two metrics so each task isolates one objective. \textbf{Construction.} For each source design, we grid-sweep a curated subset of OpenROAD-flow-scripts variables (e.g., \texttt{ABC\_AREA}, \texttt{ASAP7\_USE\_VT}, \texttt{CORE\_UTILIZATION}, \texttt{PLACE\_DENSITY}, \texttt{SYNTH\_HIERARCHICAL}; full list in \cref{app:ppa-knobs}) and identify the configuration with the best target-metric value. This configuration defines the reference fix and task target; task instances are then created by controlled perturbations that move the design \emph{away} from the reference. \textbf{L1--L2 (single- vs. multi-knob perturbation).} \textbf{L1} perturbs one uniformly sampled knob to another reasonable value that worsens the target metric. \textbf{L2} jointly perturbs several knobs, typically $2$--$3$, under the same degradation constraint. The agent is \emph{not} told which knobs were changed; it observes only the degraded post-flow state and must localize responsible parameters through tool calls. \textbf{L3 (Performance-only re-architecture).} Some performance gaps cannot be closed by knob retuning and require source-level restructuring, such as pipelining, retiming, or SDC changes. We restrict L3 to Performance because Power and Area gaps in our pool are generally configuration-closable.

\subsubsection{PPA-Multi}
\begin{figure}[tbp]
    \centering
    \begin{subfigure}[b]{0.49\columnwidth}
        \centering
        \includegraphics[width=\linewidth]{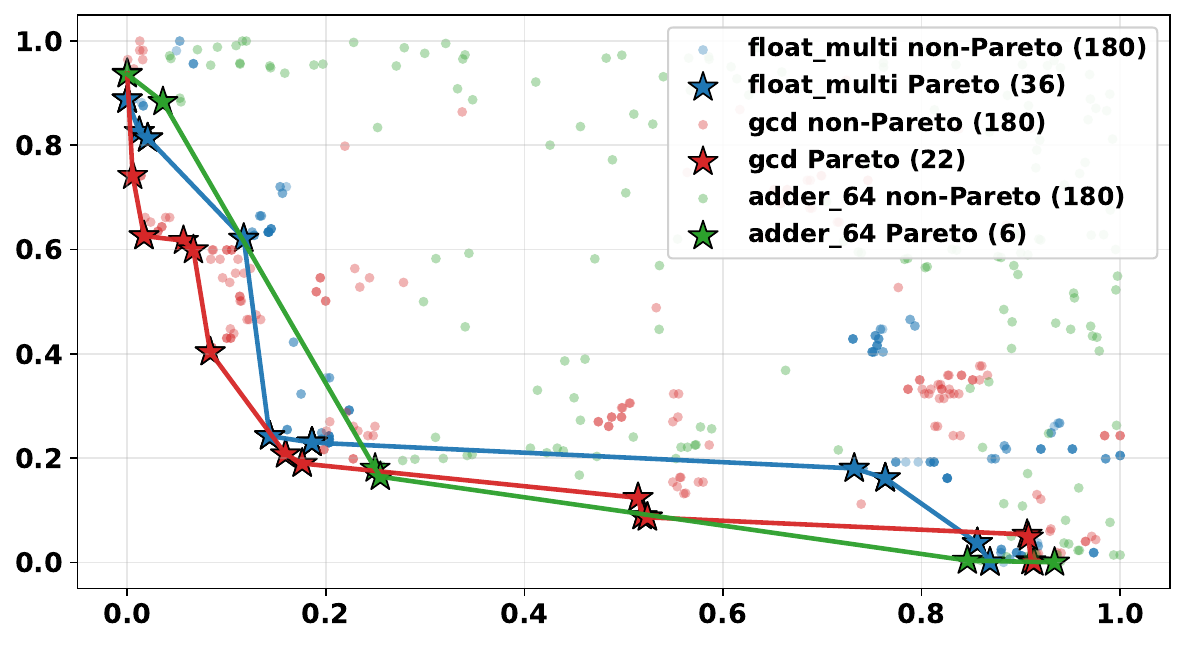}
        \caption{Effective period vs.\ total power.}
        \label{fig:pareto-combined-pp}
    \end{subfigure}
    \hfill
    \begin{subfigure}[b]{0.49\columnwidth}
        \centering
        \includegraphics[width=\linewidth]{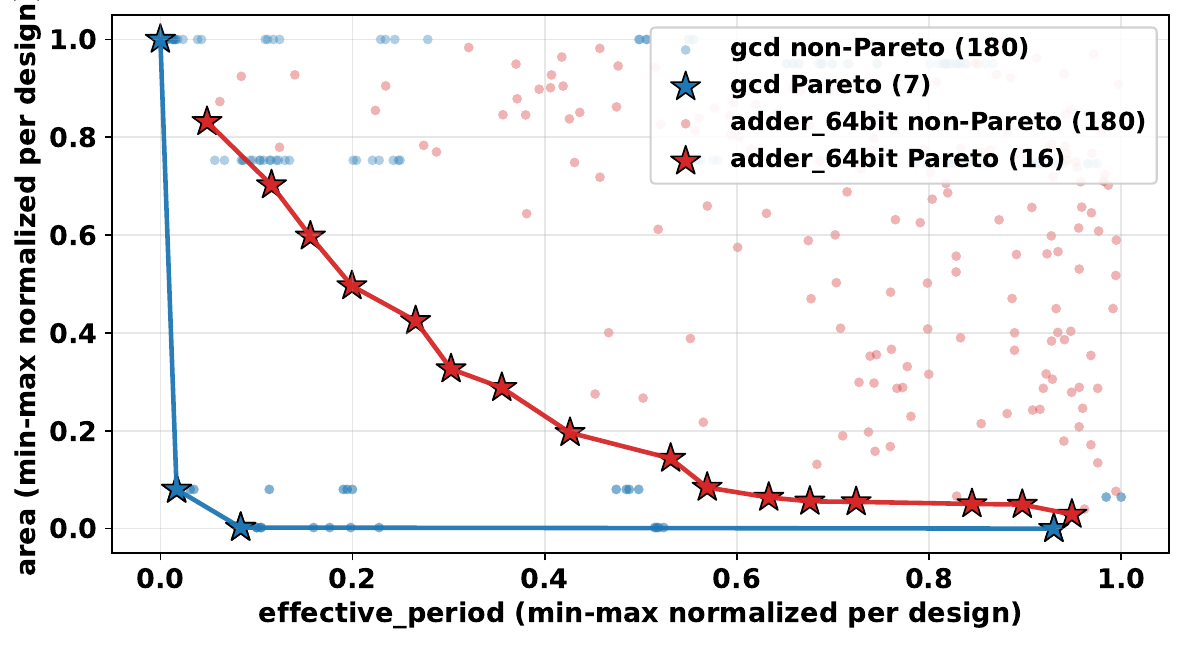}
        \caption{Effective period vs.\ die area.}
        \label{fig:pareto-combined-pa}
    \end{subfigure}
    \caption{\textbf{Two-objective Pareto fronts for representative PPA-Multi designs.} Gray dots denote non-Pareto solutions; blue stars indicate Pareto-optimal points. The combined panels summarize representative period--power and period--area trade-off fronts.}
    \label{fig:pareto}
\end{figure}
PPA-Multi evaluates multi-objective tuning under explicit metric trade-offs (light-yellow block in \cref{fig:workflow}). For each design we conduct the same grid sweep used in PPA-Mono, and for each task we set targets at points selected from the Pareto frontier so that they are jointly achievable but inaccessible by greedy single-metric optimization. \textbf{L1 (two-objective trade-off).} Each task targets a point on a 2D Pareto front. We cover all three metric pairs---performance--power, performance--area, and power--area---so the corpus jointly stresses every pairwise trade-off (\cref{fig:pareto} illustrates representative combined 2D fronts). \textbf{L2 (three-objective trade-off).} Each task targets a point on the full 3D Pareto front spanning performance, power, and area; targets are chosen from the frontier to ensure achievability while precluding shortcut solutions. In both levels the initial configuration sits in the off-frontier interior, so the agent must \emph{reach the front from the interior} while satisfying floor constraints on all non-target metrics: the score in Eq.\,\ref{eq:score-ppa} explicitly penalizes regressions on those constraints. PPA-Multi therefore tests whether agents can reason about coupled parameters and balance non-trivial trade-offs, rather than greedily optimize a single dimension.
\begin{table*}[]
\centering
\small
\caption{\textbf{Overall performance on \textsc{PostEDA-Bench}.} For both SR and VRR/NIS, all values are in \% and \emph{larger is better}. Within each performance column, the largest observed mean is \textbf{bold} and the second-largest is \underline{underlined}; ties share formatting. These markings do not imply statistical significance (\cref{app:variance}). ``--'' indicates that the framework does not apply to that dimension (ORFS-Agent targets PPA only). Measured API charges and elapsed evaluation times are reported in \cref{tab:runtime-cost}.}
\label{tab:overall-performance}
\resizebox{\textwidth}{!}{%
\begin{tabular}{
lcccccccc
}
\hline
 & \multicolumn{2}{c}{\cellcolor[HTML]{F2C6AB}\textbf{DRC-Essential}}
 & \multicolumn{2}{c}{\cellcolor[HTML]{CFEBF9}\textbf{DRC-Reasoning}}
 & \multicolumn{2}{c}{\cellcolor[HTML]{BFE5A8}\textbf{PPA-Mono}}
 & \multicolumn{2}{c}{\cellcolor[HTML]{F7E39E}\textbf{PPA-Multi}} \\ \cline{2-9}
\multirow{-2}{*}{} &
\multicolumn{1}{c}{\cellcolor[HTML]{F8E2D5}\textit{SR(\%)}} &
\multicolumn{1}{c}{\cellcolor[HTML]{F8E2D5}\textit{VRR(\%)}} &
\multicolumn{1}{c}{\cellcolor[HTML]{E1F3FD}\textit{SR(\%)}} &
\multicolumn{1}{c}{\cellcolor[HTML]{E1F3FD}\textit{VRR(\%)}} &
\multicolumn{1}{c}{\cellcolor[HTML]{DAF2CE}\textit{SR(\%)}} &
\multicolumn{1}{c}{\cellcolor[HTML]{DAF2CE}\textit{NIS(\%)}} &
\multicolumn{1}{c}{\cellcolor[HTML]{FEF2C8}\textit{SR(\%)}} &
\multicolumn{1}{c}{\cellcolor[HTML]{FEF2C8}\textit{NIS(\%)}} \\ \hline

\multicolumn{9}{c}{\cellcolor[HTML]{C0C0C0}\textit{React}} \\
GPT-5          & 84.50 & 91.94 & 31.33 & 33.80 & 39.42 & 51.59 & 11.50 & 23.69 \\
GPT-5-mini     & 65.00 & 75.21  & 20.66 & 21.88 & 42.28 & 47.74 & 10.00 & 17.36 \\
Gemini-3-Flash-preview & \underline{85.50} & 87.78 & \underline{36.66} & \underline{36.75} & \underline{48.56} & \underline{54.16} & \underline{18.50} & \underline{51.62} \\
Claude Opus 5 & \textbf{95.00} & \textbf{95.17} & \textbf{54.67} & \textbf{54.67} & 40.57 & 47.69 & 2.00 & -1.39 \\
\hdashline

DeepSeek-V3.2 & 71.00 & 79.78 & 16.66 & 18.64 & 1.14 & 7.84 &0.00  &  1.33\\
Qwen3.5-122B-A10B    & 51.50 & 58.97 & 13.33 & 14.09 & 39.42 & 47.84 & 4.00 & 35.46 \\
Gemma-4-31B-it    & 82.00 & \underline{92.02} & 27.99 & 32.05 & \textbf{64.56} & \textbf{69.34} & 11.00 & 16.34 \\
Qwen3.5-27B    & 68.50 & 77.66 & 27.32 & 30.07 & 30.85 & 42.58 & 0.0 & -39.34 \\
Qwen3.5-9B  & 56.00 & 64.85 & 5.99 & 5.99 & 25.71 & 36.56 & 1.00  & 10.88 \\
\hline
\multicolumn{9}{c}{\cellcolor[HTML]{C0C0C0}\textit{Proposer-Critic}}\\
GPT-5          & 60.50 & 71.57 & 15.99 & 17.00 & 42.85 & 50.36 & 0.50 & 11.79 \\
\hdashline

Qwen3.5-122B-A10B    &66.50  & 75.27 & 17.33 & 18.89 & 39.99 & \underline{54.16} & 2.50 & 25.03 \\
Gemma-4-31B-it    & 78.00 & 87.00 & 31.33 & 33.51 & 42.28 & 49.95  & 2.00 & -1.33 \\
Qwen3.5-9B  & 54.50 & 62.58 & 8.66 & 9.18 & 3.99 & 13.61 & 0.00 & -16.54 \\
\hline

\hline
\multicolumn{9}{c}{\cellcolor[HTML]{C0C0C0}\textit{Domain-specific Agent (ORFS-agent$^\ddagger$)}}\\
GPT-5          & -- & -- & -- & -- & 25.71 & 33.75 & 5.00 & 50.28 \\
\hdashline

Qwen3.5-122B-A10B   & -- & -- & -- & -- & 40.00 & 48.91 & \textbf{20.00} & \textbf{67.80} \\
Gemma-4-31B-it    & -- & -- & -- & -- & 47.14 & 47.31 & 1.50 & -15.90 \\
\hline
\multicolumn{9}{l}{$\S$: We exclude other open-sourced ChatEDA\cite{wu2024chateda} and MCP4EDA\cite{wang2025mcp4edallmpoweredmodelcontext} that generates scripts from prompts.}\\
\multicolumn{9}{l}{$\ddagger$: We automate ORFS-Agent search-space inference by removing manual parameter/range definitions.
}
\end{tabular}%
}
\end{table*}

\section{Experiments}
\label{sec:experiments}

\subsection{Experimental Setup}
\label{sec:exp-setup}

\textbf{Models.} We evaluate four commercial models (GPT-5, GPT-5-mini, Gemini-3-Flash-preview, Claude Opus 5) and five open-source models spanning MoE (DeepSeek-V3.2, Qwen3.5-122B-A10B), mid-size dense (Qwen3.5-27B, Gemma-4-31B-it), and small dense (Qwen3.5-9B). \textbf{Agent frameworks.} We compare \textbf{ReAct}~\cite{yao2023react}, \textbf{Proposer--Critic}~\cite{madaan2023selfrefine}, and an adapted \textbf{ORFS-Agent}~\cite{ghose2025orfsagenttoolusingagentschip} (Bayesian optimization, PPA only; see \cref{app:agent-frameworks}). \textbf{Run protocol.} Each evaluated (model, scaffold, task) combination runs $5$ times; we report mean SR/VRR/NIS. Temperature is $0$ where supported; Claude Opus 5 rejects this API parameter, so it is omitted for Claude. Paired task-level bootstrap CIs appear in \cref{app:variance}. ReAct and Proposer--Critic use adaptive caps of $8$ DRC / $16$ PPA agent actions; ORFS-Agent uses $18$ OpenROAD runs per PPA task (\cref{app:iteration-semantics,app:exp-protocol}).

\subsection{Main Results}
\label{sec:main-results}
\Cref{tab:overall-performance} shows that no evaluated configuration leads all four task families. Claude Opus 5 has the largest observed DRC-Essential SR/VRR ($95.00\%/95.17\%$) and DRC-Reasoning SR/VRR ($54.67\%/54.67\%$); Gemma-4-31B-it leads PPA-Mono SR/NIS ($64.56\%/69.34\%$), and ORFS-wrapped Qwen3.5-122B leads PPA-Multi ($20.00\%/67.80\%$). DeepSeek-V3.2 drops from $71.00\%$ DRC-Essential SR to near-zero PPA. \textbf{The synthetic-to-practical gap persists with Claude:} the best DRC-Reasoning SR is only $\sim 58\%$ of the best DRC-Essential SR, while PPA-Multi reaches only $\sim 31\%$ of PPA-Mono SR. \textbf{Geometric reasoning is decoupled from rule recall:} GPT-5, Gemini-3-Flash, Gemma-4, and Claude retain approximately $34$--$58\%$ of their DRC-Essential SR on DRC-Reasoning, so the post-flow repair gap remains despite improved backbone capability. \textbf{PPA-Mono is target-shortfall:} agents move the right metrics ($\mathrm{NIS}>\mathrm{SR}$ by $5$--$12$ points) but stop short of the target. \textbf{PPA-Multi is trade-off failure:} negative NIS (e.g., Qwen3.5-27B $-39.34$; Gemma-4/ORFS $-15.90$; Claude $-1.39$) reflects constraint regression while chasing the violated objective rather than balancing competing targets. \textbf{Scaffold effects are dimension-conditioned.} Proposer--Critic improves some DRC cells by validating edits before dispatch but hurts PPA by over-constraining exploration. ORFS-Agent is complementary: Bayesian optimization lifts Qwen3.5-122B on PPA-Multi ($4.00/35.46\!\to\!20.00/67.80$ SR/NIS) yet amplifies weak inferred search spaces for Gemma-4 ($16.34\!\to\!-15.90$ NIS). Framework choice is thus not a universal upgrade; it trades off edit validation, search-policy flexibility, and search-space coverage (\cref{app:extended-results}).

\subsubsection{DRC-Bench: per-level breakdown}
\label{sec:drc-perlevel}
To localize \emph{which} DRC sub-skill fails, \cref{tab:drc-bench} decomposes SR and VRR over construction-time levels for three representative ReAct backbones. Parenthesized $\Delta$ values report the difference from that model's overall score in \cref{tab:overall-performance}. \textbf{Essential L1/L2 are saturated; L3 is the discriminator.} All three backbones reach $92$--$96\%$ SR on Essential L1/L2, so atomic rule recall and contextual clutter filtering are largely solved. The L3 cascade drops to $44$--$62\%$ SR, while VRR remains much higher for GPT-5 and Gemma-4 ($54.00/82.78$, $44.00/84.09$ SR/VRR), showing that agents often fix most violations but leave one residual after interacting edits. This localizes the bottleneck to edit-interaction modeling rather than local rule knowledge. \textbf{Reasoning collapses after the first step.} On DRC-Reasoning, L1 remains partially solvable ($56.91$--$63.07\%$ SR), but L2/L3 fall to $2.00$--$25.71\%$ SR. Gemini-3's advantage is concentrated exactly in the multi-step regimes (L2/L3: $18.00/25.71$ vs GPT-5 $6.00/8.56$ and Gemma-4 $2.00/8.56$), not on L1. Thus practical post-flow violations primarily stress trajectory coherence and coupled geometric planning, not single-edit accuracy.
\begin{table}[t]
\centering
\small
\caption{\textbf{DRC-Bench per-level performance.} SR/VRR for representative ReAct backbones on L1--L3; parenthesized $\Delta$ values report the difference from the corresponding overall DRC score. The best score in each column is bolded, with ties sharing bold.}
\label{tab:drc-bench}
\setlength{\tabcolsep}{2.2pt}
\renewcommand{\arraystretch}{1.15}
\resizebox{\columnwidth}{!}{%
\begin{tabular}{lcccccc}
\hline
 & \multicolumn{2}{c}{\textbf{L1}} & \multicolumn{2}{c}{\textbf{L2}} & \multicolumn{2}{c}{\textbf{L3}} \\ \cline{2-7}
 & \textit{SR(\%)} & \textit{VRR(\%)} & \textit{SR(\%)} & \textit{VRR(\%)} & \textit{SR(\%)} & \textit{VRR(\%)} \\ \hline
\multicolumn{7}{c}{\cellcolor[HTML]{F2C6AB}\textit{DRC-Essential}} \\
GPT-5 & \textbf{94.00} ($+9.50$) & \textbf{94.00} ($+2.06$) & \textbf{96.00} ($+11.50$) & \textbf{97.00} ($+5.06$) & 54.00 (\textcolor{red}{$-30.50$}) & 82.78 (\textcolor{red}{$-9.16$}) \\
Gemini-3-Flash & 92.00 ($+6.50$) & 92.00 ($+4.22$) & \textbf{96.00} ($+10.50$) & 96.00 ($+8.22$) & \textbf{62.00} (\textcolor{red}{$-23.50$}) & 71.13 (\textcolor{red}{$-16.65$}) \\ \hdashline
Gemma-4-31B-it & \textbf{94.00} ($+12.00$) & \textbf{94.00} ($+1.98$) & \textbf{96.00} ($+14.00$) & 96.00 ($+3.98$) & 44.00 (\textcolor{red}{$-38.00$}) & \textbf{84.09} (\textcolor{red}{$-7.93$}) \\
\hline
\multicolumn{7}{c}{\cellcolor[HTML]{CFEBF9}\textit{DRC-Reasoning}} \\
GPT-5 & \textbf{63.07} ($+31.74$) & \textbf{63.07} ($+29.27$) & 6.00 (\textcolor{red}{$-25.33$}) & 7.00 (\textcolor{red}{$-26.80$}) & 8.56 (\textcolor{red}{$-22.77$}) & 17.74 (\textcolor{red}{$-16.06$}) \\
Gemini-3-Flash & 56.91 ($+20.25$) & 56.91 ($+20.16$) & \textbf{18.00} (\textcolor{red}{$-18.66$}) & \textbf{18.00} (\textcolor{red}{$-18.75$}) & \textbf{25.71} (\textcolor{red}{$-10.95$}) & \textbf{26.12} (\textcolor{red}{$-10.63$}) \\ \hdashline
Gemma-4-31B-it & 58.45 ($+30.46$) & 58.45 ($+26.40$) & 2.00 (\textcolor{red}{$-25.99$}) & 2.00 (\textcolor{red}{$-30.05$}) & 8.56 (\textcolor{red}{$-19.43$}) & 25.95 (\textcolor{red}{$-6.10$}) \\
\hline
\multicolumn{7}{l}{$\Delta$: Difference from the corresponding overall DRC score in \cref{tab:overall-performance}. $\dagger$: All rows use the ReAct workflow.}\\
\end{tabular}%
}
\end{table}



\subsubsection{PPA-Bench: per-level breakdown}
\label{sec:ppa-perlevel}
We focus the main paper on PPA-Performance and PPA-Multi, the two slices that expose re-architecture and multi-objective trade-offs. \Cref{tab:ppa-bench} reports SR/NIS for Gemini-3-Flash and Gemma-4-31B-it under ReAct, plus ORFS+Qwen-122B under ORFS-Agent; PPA-Power/Area and the full bar plots are deferred to \cref{app:ppa-perlevel}. \textbf{PPA-Performance L3 is the re-architecture wall.} L3 drops to only $8$--$20\%$ SR and $20$--$27\%$ NIS across all three backbones, with negative deltas for every L3 cell. This isolates RTL micro-architectural reasoning as the bottleneck: agents can improve shallow timing knobs on L1/L2, but fail when retiming or pipelining is structurally required. \textbf{PPA-Multi exposes constraint-tracking failure, while ORFS improves coverage.} Moving from L1 (two-objective) to L2 (three-objective) lowers SR/NIS for every backbone; Gemma's L2 NIS even turns negative ($-0.10\%$), indicating that as the number of competing objectives grows, agents increasingly regress constraint metrics rather than balance the trade-off---the same trade-off bottleneck flagged in the overall PPA-Multi results. ORFS+Qwen-122B still leads PPA-Multi overall ($20.00/67.80$ SR/NIS), so structured exploration helps cover the Pareto frontier, but it does not remove the L2 trade-off gap.

\begin{table}[t]
\centering
\small
\caption{\textbf{PPA-Bench per-level performance.} SR/NIS for representative backbones on PPA-Performance and PPA-Multi; parenthesized $\Delta$ values report the difference from the corresponding dimension-level overall score. The best score in each column is bolded, with ties sharing bold.}
\label{tab:ppa-bench}
\setlength{\tabcolsep}{2.2pt}
\renewcommand{\arraystretch}{1.15}
\resizebox{\columnwidth}{!}{%
\begin{tabular}{lcccccc}
\hline
\multicolumn{7}{c}{\cellcolor[HTML]{BFE5A8}\textit{PPA-Performance (Mono)}} \\
 & \multicolumn{2}{c}{\textbf{L1}} & \multicolumn{2}{c}{\textbf{L2}} & \multicolumn{2}{c}{\textbf{L3}} \\ \cline{2-7}
 & \textit{SR(\%)} & \textit{NIS(\%)} & \textit{SR(\%)} & \textit{NIS(\%)} & \textit{SR(\%)} & \textit{NIS(\%)} \\ \hline
Gemini-3-Flash & 56.00 ($+21.34$) & 59.99 ($+17.70$) & \textbf{40.00} ($+5.34$) & 40.28 (\textcolor{red}{$-2.01$}) & 8.00 (\textcolor{red}{$-26.66$}) & \textbf{26.59} (\textcolor{red}{$-15.70$}) \\ \hdashline
Gemma-4-31B-it & \textbf{84.00} ($+37.34$) & \textbf{84.00} ($+34.58$) & 36.00 (\textcolor{red}{$-10.66$}) & 37.78 (\textcolor{red}{$-11.64$}) & \textbf{20.00} (\textcolor{red}{$-26.66$}) & 26.50 (\textcolor{red}{$-22.92$}) \\
ORFS+Qwen-122B & 20.00 (\textcolor{red}{$-6.66$}) & 38.61 ($+3.27$) & \textbf{40.00} ($+13.34$) & \textbf{47.40} ($+12.06$) & \textbf{20.00} (\textcolor{red}{$-6.66$}) & 20.00 (\textcolor{red}{$-15.34$}) \\
\hline
\multicolumn{7}{c}{\cellcolor[HTML]{F7E39E}\textit{PPA-Multi}} \\
 & \multicolumn{2}{c}{\textbf{L1}} & \multicolumn{2}{c}{\textbf{L2}} & \multicolumn{2}{c}{\textbf{Overall}} \\ \cline{2-7}
 & \textit{SR(\%)} & \textit{NIS(\%)} & \textit{SR(\%)} & \textit{NIS(\%)} & \textit{SR(\%)} & \textit{NIS(\%)} \\ \hline
Gemini-3-Flash & \textbf{25.00} ($+6.50$) & 54.93 ($+3.31$) & 12.00 (\textcolor{red}{$-6.50$}) & 48.31 (\textcolor{red}{$-3.31$}) & 18.50 ($+0.00$) & 51.62 ($+0.00$) \\ \hdashline
Gemma-4-31B-it & 17.00 ($+6.00$) & 32.77 ($+16.43$) & 5.00 (\textcolor{red}{$-6.00$}) & -0.10 (\textcolor{red}{$-16.44$}) & 11.00 ($+0.00$) & 16.34 ($+0.00$) \\
ORFS+Qwen-122B & \textbf{25.00} ($+5.00$) & \textbf{73.85} ($+6.05$) & \textbf{15.00} (\textcolor{red}{$-5.00$}) & \textbf{61.75} (\textcolor{red}{$-6.05$}) & \textbf{20.00} ($+0.00$) & \textbf{67.80} ($+0.00$) \\
\hline
\multicolumn{7}{l}{$\Delta$: Difference from the corresponding dimension-level overall score; negative deltas are red.}\\
\multicolumn{7}{l}{$\dagger$: Gemini-3-Flash and Gemma-4-31B-it use ReAct; ORFS+Qwen-122B uses ORFS-Agent.}\\
\multicolumn{7}{l}{$\ddagger$: PPA-Power/Area figures and analysis are provided in \cref{app:ppa-perlevel}.}
\end{tabular}%
}
\end{table}

\begin{figure}[bt]
  \centering
  \begin{subfigure}[b]{0.49\columnwidth}
    \centering
    \includegraphics[width=\linewidth]{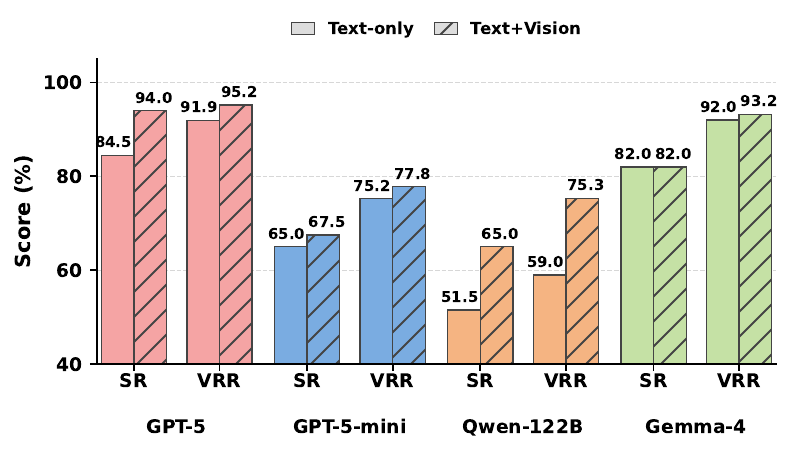}
    \caption{DRC-Essential.}
    \label{fig:drc-vision-essential}
  \end{subfigure}
  \hfill
  \begin{subfigure}[b]{0.49\columnwidth}
    \centering
    \includegraphics[width=\linewidth]{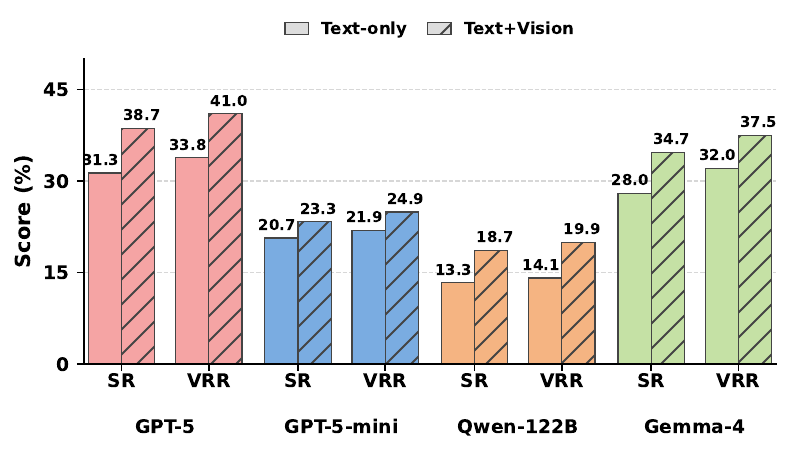}
    \caption{DRC-Reasoning.}
    \label{fig:drc-vision-reasoning}
  \end{subfigure}
  \caption{\textbf{Effect of vision modality on DRC-Bench.} SR and VRR are combined in each subfigure for four backbones under text-only and text+vision settings.}
  \label{fig:drc-vision}
\end{figure}
\subsection{Effect of Vision Modality on DRC-Bench}
\label{sec:vision-ablation}
We pair each text-only backbone (GPT-5, GPT-5-mini, Qwen3.5-122B-A10B, Gemma-4-31B-it) with its vision-augmented variant under ReAct; vision runs add \texttt{vision\_query\_with\_pts}, which renders a KLayout image of the violation region (\cref{fig:drc-vision}). \textbf{Vision augmentation enhances DRC-Bench:} across all $16$ observed (model, dimension, metric) cells the layout-image channel does not reduce SR or VRR (the only zero-delta is Gemma's DRC-Essential SR, $82.00\!\to\!82.00$) and yields consistent gains on both DRC-Essential and DRC-Reasoning, supplying geometric evidence that text-only rule identifiers and coordinates do not carry. \textbf{DRC-Essential gains vary across backbones.} Qwen3.5-122B has the largest observed gains ($+13.5$ SR and $+16.3$ VRR) from a low baseline ($51.50/58.97$). GPT-5 also gains $+9.5$ SR ($84.50\!\to\!94.00$), whereas Gemma-4's SR is unchanged and its VRR increases by $+1.20$. \textbf{DRC-Reasoning gains are positive across backbones.} SR gains are positive for every backbone ($+2.67$ to $+7.33$), with similar VRR gains ($+3.04$ to $+7.22$); GPT-5-mini gains $+2.67$ SR and $+3.04$ VRR, compared with GPT-5's $+7.33$ SR and $+7.22$ VRR. These cell-wise gains are descriptive: only two of eight individual SR confidence intervals exclude zero, while the pooled SR gains are positive for both task families (\cref{app:variance}).


\subsection{Effect of Iteration Budget}
\label{sec:iter-ablation}
\begin{figure*}[t]
  \centering
  \includegraphics[width=0.95\textwidth]{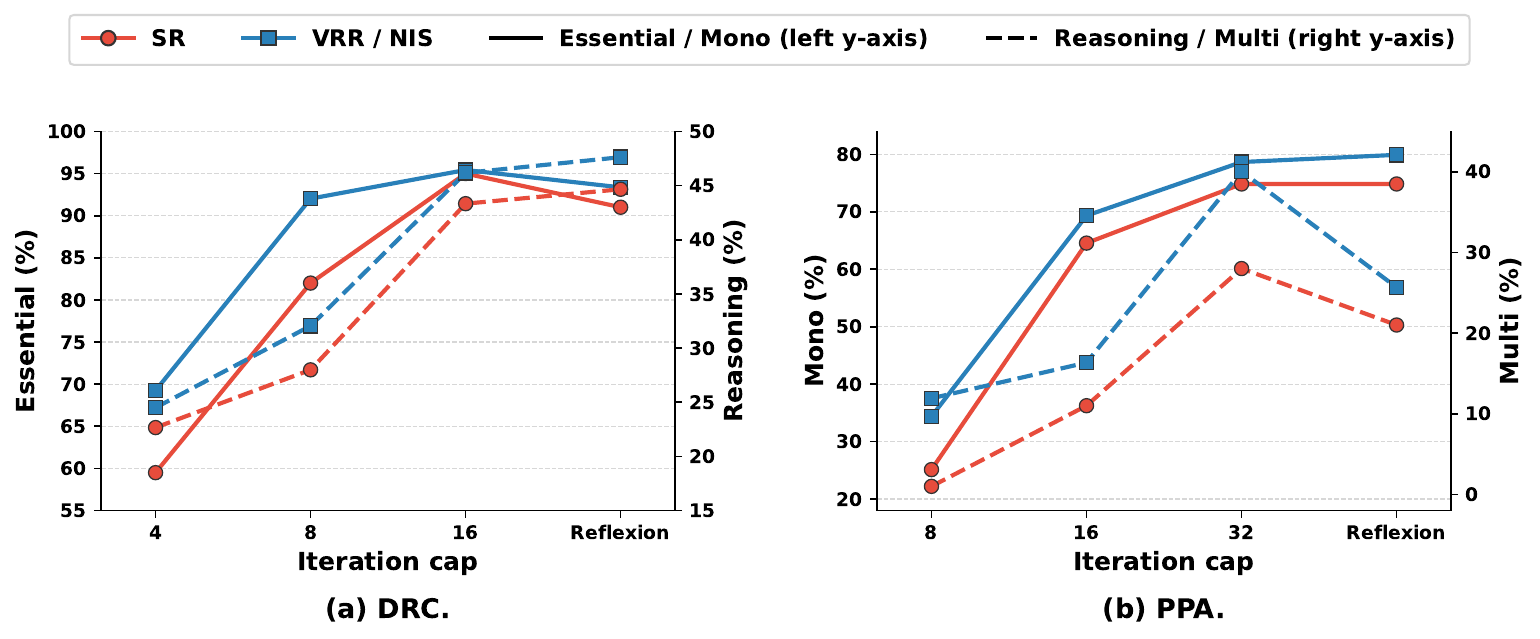}
  \caption{\textbf{Effect of iteration cap and Reflexion on Gemma-4-31B-it.} Left: DRC combines DRC-Essential and DRC-Reasoning. Right: PPA combines PPA-Mono and PPA-Multi. Colors encode metrics; solid lines use the left y-axis and dashed lines use the right y-axis.}
  \label{fig:iter-ablation}
  \vspace{-0.8em}
\end{figure*}

We sweep the iteration cap and compare \textbf{Reflexion}~\cite{shinn2023reflexion} on Gemma-4-31B-it under ReAct (\cref{fig:iter-ablation}). \textbf{Reflexion protocol.} Reflexion runs two attempts of the per-attempt cap ($8$ for DRC, $16$ for PPA). The second attempt starts \emph{completely fresh} from the original task state---no GDS edits, no flow artifacts, only a verbal reflection summarized from attempt~1 are carried over (\cref{app:reflexion-tot}). The fair baseline is therefore ReAct at the same per-attempt cap, since each Reflexion trial sees the same per-trial budget; any improvement over that baseline at twice the total cost reflects genuine value from verbal reinforcement, not extra iterations within a single trajectory. \textbf{Iteration scaling is dimension-specific.} DRC-Essential and PPA-Mono show diminishing returns, while DRC-Reasoning and PPA-Multi are still climbing at the largest measured caps---one cap saturates easy tasks while leaving hard ones budget-sensitive. \textbf{Verbal reflection improves on the per-attempt baseline.} At the per-attempt cap (iter=$8$/$16$), Reflexion lifts DRC-Essential SR from $82.00$ to $91.00$, DRC-Reasoning from $27.99$ to $44.66$, PPA-Mono from $64.56$ to $74.85$, and PPA-Multi from $11.00$ to $21.00$, so the carried-over reflection truly transfers between independent trials.

\subsection{Effect of Thinking Mode}
\label{sec:thinking-ablation}
\begin{table}[tb]
\centering
\caption{\textbf{Thinking-mode ablation} for Gemma-4-31B-it under ReAct.}
\label{tab:thinking}
\resizebox{\columnwidth}{!}{%
\begin{tabular}{lccccccccc}
\hline
\multirow{2}{*}{\textbf{}} 
& \multirow{2}{*}{\textbf{Thinking}} 
& \multicolumn{2}{c}{\cellcolor[HTML]{F2C6AB}\textbf{DRC-Essential}} 
& \multicolumn{2}{c}{\cellcolor[HTML]{CFEBF9}\textbf{DRC-Reasoning}} 
& \multicolumn{2}{c}{\cellcolor[HTML]{BFE5A8}\textbf{PPA-Mono}} 
& \multicolumn{2}{c}{\cellcolor[HTML]{F7E39E}\textbf{PPA-Multi}} \\ \cline{3-10}

& 
& \multicolumn{1}{c}{\cellcolor[HTML]{F8E2D5}\textit{SR(\%)}} 
& \multicolumn{1}{c}{\cellcolor[HTML]{F8E2D5}\textit{VRR(\%)}} 
& \multicolumn{1}{c}{\cellcolor[HTML]{E1F3FD}\textit{SR(\%)}}
& \multicolumn{1}{c}{\cellcolor[HTML]{E1F3FD}\textit{VRR(\%)}}
& \multicolumn{1}{c}{\cellcolor[HTML]{DAF2CE}\textit{SR(\%)}}
& \multicolumn{1}{c}{\cellcolor[HTML]{DAF2CE}\textit{NIS(\%)}}
& \multicolumn{1}{c}{\cellcolor[HTML]{FEF2C8}\textit{SR(\%)}} 
& \multicolumn{1}{c}{\cellcolor[HTML]{FEF2C8}\textit{NIS(\%)}} \\ \hline

\multicolumn{10}{c}{\cellcolor[HTML]{C0C0C0}\textit{React}} \\

\multirow{3}{*}{Gemma-4-31B-it} 
& On  & 82.00 & \textbf{92.02} & 27.99 & \textbf{32.05} & \textbf{64.56} & \textbf{69.34} & 11.00 & 16.34 \\
& Off & 51.00 & 63.75 & 8.66 & 8.66 & 54.28 & 65.20 & \textbf{18.50} & \textbf{38.40} \\
& ToT & \textbf{85.50} & 90.35 & \textbf{29.99} & 31.55 & 31.42 & 40.53 & 0.00 & 1.35 \\ \hline

\end{tabular}%
}
\vspace{-0.8em}
\end{table}
We compare standard chain-of-thought (\textsc{On}), thinking suppressed (\textsc{Off}), and Tree-of-Thought (\textsc{ToT})~\cite{yao2023tot} (\cref{app:reflexion-tot}) on Gemma-4-31B-it under ReAct (\cref{tab:thinking}). \textbf{Thinking is dimension-conditioned.} Turning thinking on improves DRC-Essential SR by $31$ points, DRC-Reasoning by $19.3$, and PPA-Mono by $10.3$, but hurts PPA-Multi (SR/NIS drop from $18.50/38.40$ to $11.00/16.34$): DRC and PPA-Mono benefit from reasoning before acting, while PPA-Multi benefits more from faster exploration and tool dispatch under a fixed iteration cap. \textbf{ToT helps DRC slightly but fails on PPA.} ToT raises DRC SR modestly ($+3.5$ Essential, $+2.0$ Reasoning) without improving VRR, suggesting branching closes a few near-complete fixes rather than opening new paths; on PPA it sharply reduces Mono ($31.42/40.53$ vs.\ $64.56/69.34$ SR/NIS) and nearly fails Multi ($0.00/1.35$), so extra deliberation is not a substitute for search coverage.

\vspace{-10pt}
\section{Conclusion}
We presented \sysname, a $145$-task hierarchical benchmark that jointly evaluates sign-off DRC fixing and PPA convergence. Experiments across nine LLMs under multiple agent scaffolds expose a synthetic-to-practical asymmetry---agents do well on DRC-Essential and PPA-Mono but degrade sharply on DRC-Reasoning and PPA-Multi---and identify trade-off reasoning, rather than knob knowledge, as the PPA-Multi bottleneck; vision augmentation improves pooled DRC success rates. \sysname provides a grounded testbed to drive future research on geometry-aware DRC repair and trade-off-faithful PPA optimization.

\section*{Impact Statements}
This paper presents work whose goal is to advance the field of machine learning. There are many potential societal consequences of our work, none of which we feel must be specifically highlighted here.

\begin{ack}
\end{ack}

\bibliographystyle{abbrvnat}
\bibliography{reference}

\appendix

\section{Additional Related Work}
\label{app:related-work}
\subsection{EDA background}
\label{app:eda-background}
\paragraph{Layout and PDK layers.}
A physical layout is a stack of two-dimensional polygon layers defined by a process design kit (PDK). These include transistor-related layers and metal interconnect layers, with vias connecting conductors across layers. In \sysname, the layout tools expose transistor-related layers, metals \texttt{M1}--\texttt{M9}, and vias \texttt{V0}--\texttt{V9} (\cref{app:tools}); the process is ASAP7~\cite{clark2016asap7}. Each layout is supplied as a GDS file, a binary geometric database containing the polygons and cell hierarchy that the agent inspects and edits.

\paragraph{DRC decks, sign-off, and residual violations.}
A DRC deck is a machine-readable implementation of the foundry or PDK design rules. For a fixed layout, deck, and checker configuration, running DRC determines which geometric constraints are violated and returns their locations. Sign-off DRC applies the designated deck to assess whether the final layout satisfies those rules. Synthesis, placement, and routing produce the layout; violations that remain after the flow and its built-in repair procedures are \emph{residuals}. Engineers inspect and repair these residuals while avoiding new violations. Such post-flow residuals are the source of DRC-Reasoning (\cref{sec:drc-reasoning}).

\paragraph{Kinds of violation.}
The rule families considered here constrain geometric relations between polygons on one or two layers: minimum width, spacing, enclosed or overlap area, and layer-pair extension. For example, a spacing rule specifies the minimum permitted separation between shapes, while an extension rule requires one layer to extend beyond another by a prescribed distance. A violation is therefore a spatial property of the actual layout. Its rule identifier and textual description specify the requirement, but determining a valid repair also requires the offending coordinates and surrounding geometry.

\paragraph{PPA constraints.}
Power--performance--area (PPA) targets constrain total power, clock period (represented by effective period in our evaluation), and die area. These quantities are coupled: a change that improves one can worsen another. Multi-objective tasks therefore combine optimization targets with quality floors on metrics that must remain acceptable; for the lower-is-better quantities used here, these floors are expressed as upper bounds. Agents must meet all specified thresholds by adjusting flow parameters, timing constraints, or RTL. PPA-Mono isolates one objective without bounds on the other two, whereas PPA-Multi explicitly evaluates simultaneous target attainment (\cref{eq:score-ppa}).

\subsection{Evaluation gaps in prior EDA-agent work}
\label{app:prior-benchmark-gaps}
\Cref{tab:benchmark-comparison} highlights several gaps in existing evaluations. \textbf{Objective and scale.} ChatEDA-bench~\cite{wu2024chateda}, ChipNeMo~\cite{liu2024chipnemodomainadaptedllmschip}, iEDA-bench~\cite{wu2025EDAid}, and AutoEDA~\cite{lu2025autoedaenablingedaflow} emphasize script generation or flow execution, providing limited evidence of convergence to explicit PPA targets. Optimization studies also have narrow coverage: GPT4AIGChip~\cite{Fu2023GPT4AIGChip}, IICPilot~\cite{jiang2024iicpilotintelligentintegratedcircuit}, and ORFS-Agent~\cite{ghose2025orfsagenttoolusingagentschip} evaluate six accelerator workloads, four designs, and six circuit--technology instances, respectively.

\textbf{Hierarchy and toolchain coverage.} Existing suites, including MCP4EDA~\cite{wang2025mcp4edallmpoweredmodelcontext}, use flat or coarse task groupings that do not isolate configuration recovery, RTL re-architecture, and coupled PPA trade-offs. Evaluations are also commonly tied to a particular flow, such as Vivado HLS or OpenROAD, limiting evidence of generalization across toolchains.

\textbf{Missing DRC repair.} None of the related benchmarks in \cref{tab:benchmark-comparison} evaluates residual sign-off DRC fixing, leaving geometric reasoning, layout-context filtering, and interacting repair sequences unmeasured. These gaps motivate \sysname's joint, hierarchical evaluation of DRC repair and target-driven PPA convergence.

\section{Benchmark Construction Documentation}
\label{app:source-design-protocol}

\paragraph{Filtering and deduplication.}
The source pool is filtered in the following order: parsing, top-module resolution, synthesis, standard-cell scale, license/release eligibility, and deduplication.
Deduplication is performed across all sources rather than within each source separately.
We first flag candidates with identical or near-identical top-module names, then compare port signatures (name, direction, and bit width), and finally compare post-synthesis gate-level structure.
Designs with more than 90\% gate-level overlap are merged, retaining the instance with more complete source files, clearer provenance, or broader downstream task utility.

\paragraph{Contamination control.}
\label{app:contamination}
The prompt is the part most likely to resemble pretraining data---public source RTL, a published ASAP7 rulebook---but it is not what the agent must produce.

\textbf{DRC-Bench.} The prompt does not stand alone: the agent is handed a GDS layout, a binary geometric database, and must localize the violation inside it and emit an edit at exact coordinates KLayout then re-checks. The rulebook says M1 corner-to-corner spacing is $20$\,nm; it does not say which of \texttt{gcd}'s polygons violates it, at which coordinate, or which repair clears it without creating a new one.

\textbf{PPA-Bench.} The absorbed RTL is the starting design; the scored object is the modification that closes on this task's targets. Memorizing the public source gives an agent the state it was already handed, not the retiming, pipelining or knob setting that reaches an unseen target---verified by an OpenROAD run plus the applicable testbench-based check (\cref{app:func-equiv}). The gap between ingested RTL and required fix is the task.

\paragraph{Reproducibility of manually constructed tasks.}
\label{app:manual-reproducibility}
Most tasks are manually constructed, and evaluation is reproduced from the released initial artifacts. Each evaluation unit includes its prompt, target or violation specification, tool environment, and scoring configuration. Users can rerun an agent from that fixed initial state and independently recompute SR, VRR, and NIS from the resulting artifacts and reports. Construction scripts are retained where applicable, alongside the source-design provenance, rule or knob choices, and task metadata documenting the manually selected cases. The DRC step-count protocol is documented separately in \cref{app:step-protocol}.

All released tasks can be evaluated and scored with open-source tools: KLayout for DRC and OpenROAD for PPA. Tool versions, per-task manifests, and evaluation drivers are specified in \cref{app:versions,app:manifest,app:ppa-reproducibility,app:ppa-manifest}. Fixed artifacts and scoring rules make outcomes auditable, while five-run evaluation and retained tool traces document agent behavior (\cref{app:models,app:variance}).

\paragraph{Release package and versioning.}
\label{app:versioning}
The release package contains task definitions, prompts, source RTL, GDS files for DRC-Bench, flow configurations, DRC decks, report parsers, scoring scripts, agent tool interfaces, and machine-readable dataset metadata describing provenance, licenses, and intended use.
For DRC-Bench, released layouts, prompts, sign-off reports, and KLayout setup are sufficient for agent evaluation without regenerating Innovus-derived tasks; for PPA-Bench, each released OpenROAD project directory includes the files, reports, driver, and pinned flow setup needed for re-evaluation.

\section{DRC-Bench: Reproducibility, Tool Contracts, and Construction Details}
\label{app:reproducibility}

\subsection{Software stack and tool versions}
\label{app:versions}
DRC-Bench is reproduced end-to-end with the following pinned stack:
\begin{itemize}[leftmargin=*]
    \item \textbf{PDK:} ASAP7~\cite{clark2016asap7}, the predictive 7\,nm process design kit; the bundled DRC deck is used as-is for sign-off.
    \item \textbf{Sign-off DRC:} \textsc{KLayout} v0.30.3 driven by the ASAP7 DRC deck. KLayout alone is sufficient to score any submitted agent.
    \item \textbf{LVS availability:} The certified ASAP7 layout-versus-schematic (LVS) deck targets \textsc{Mentor Calibre} v2018, a legacy version no longer broadly accessible to academic groups. The original evaluation therefore uses DRC-clean scoring without a mandatory Calibre LVS gate. We subsequently wrote a KLayout-based checker to audit connectivity and device preservation in archived repairs (\cref{app:drc-functional-audit}).
    \item \textbf{Open-source flow (used to seed DRC-Reasoning L1/L2 and the post-flow layout of DRC-Essential L2):} \textsc{OpenROAD-flow-scripts}; the sweep covers the broad set of tunable variables documented in \texttt{FlowVariables.md} (clock period, utilization, placement effort, routing effort, and many others) rather than a fixed shortlist, so the resulting violation pool is diverse along multiple axes.
    \item \textbf{Commercial flow:} \textsc{Synopsys Design Compiler} W-2024.09-SP5-3 followed by \textsc{Cadence Innovus} 21.1, with sweeps over the standard synthesis-, placement-, and routing-effort knobs exposed by both tools.
    \item \textbf{Eval driver:} \texttt{eval/eval\_drc\_sr.sh}. Default configuration: \texttt{TEMPERATURE=0}, \texttt{NUM\_RUNS=5}, \texttt{MAX\_PARALLEL\_AGENTS=5}. The per-task tool-call iteration limit is treated as an evaluation hyperparameter rather than a fixed cap and is varied in our iteration ablation. The driver reports per-task SR (mean over 5 runs) and per-task VRR, aggregated per-class and overall.
    \item \textbf{Agent tools:} released under \texttt{agents/drc/tools/} (full schemas in \cref{app:tools}).
\end{itemize}

\subsection{Per-task layout and release manifest}
\label{app:manifest}
Every task is a self-contained directory:
\begin{itemize}[leftmargin=*]
    \item \texttt{top\_cell.gds}: the input layout supplied to the agent.
    \item \texttt{prompt.txt}: the natural-language prompt embedding the violation list (rule, geometry type, coordinates).
    \item \texttt{info.json}: structured metadata---rule identifier(s), initial violation count, top-cell name, and sign-off DRC runtime.
    \item \texttt{6\_drc\_count.rpt}: the raw KLayout DRC report used to construct \texttt{prompt.txt}.
    \item \texttt{create\_errors.py}, \texttt{drc\_error\_collection.py}: the construction scripts that produced the violation, retained for auditability.
\end{itemize}
Tasks are organized as \texttt{benchmark/drc\_bench/drc\_essential/\{L1,L2,L3\}/} and \texttt{benchmark/drc\_bench/drc\_reasoning/\{L1,L2,L3\}/}. The full benchmark and agent tools are released under a permissive license (CC BY 4.0) at \url{https://github.com/pengjas/posteda-bench}; per-task evaluation logs will also be released.

\subsection{Agent tool contracts}
\label{app:tools}
The five tools exposed to the agent have the following contracts (full Python signatures and return schemas are released under \texttt{agents/drc/tools/}).
\begin{itemize}[leftmargin=*]
    \item \texttt{query\_with\_pts(input\_gds, level, cell\_name, type, direct\_layers, context\_layers, query\_pts, second\_level\_index\_range, third\_level\_index\_range)}: hierarchical textual query at three zoom levels---\emph{first} (shapes directly causing the violation), \emph{second} (shapes inside the bounding box of those shapes), \emph{third} (nearest neighbours just outside the bounding box). \texttt{type} is one of \texttt{edge-pair}, \texttt{polygon}, \texttt{edge}; \texttt{query\_pts} carries coordinates in the layout's native units. The 38 valid layer names span \texttt{Active}, \texttt{Gate}, \texttt{LIG}, \texttt{LISD}, \texttt{NSELECT}, \texttt{PSELECT}, \texttt{SDT}, \texttt{M1}--\texttt{M9}, \texttt{V0}--\texttt{V9}, and several PDK-specific markers. Returned shapes are sorted by proximity to the violation; pagination over context shapes is exposed via \texttt{*\_index\_range}.
    \item \texttt{vision\_query\_with\_pts(\dots)}: same query semantics as above but renders a KLayout image of the queried region. The image is converted to RGB JPEG, thumbnail-resized to fit within $1024{\times}1024$ at quality 85, Base64-encoded, and returned as a data URI for direct ingestion by vision-language models.
    \item \texttt{add\_shape(\dots)}, \texttt{change\_shape(\dots)}, \texttt{move\_cell(\dots)}: editing primitives that operate on the working GDS in place; they validate layer membership and shape primitives and reject illegal edits with structured error messages.
    \item \texttt{query\_drc\_report(\dots)}: re-runs the KLayout sign-off DRC deck on the current GDS and returns the updated violation list. Each call counts as one iteration against the per-task iteration limit (varied across experiments).
\end{itemize}

\subsection{DRC-Essential rule and layer coverage}
\label{app:rule-coverage}
DRC-Essential L1 and L2 jointly cover several rule families and 14 layers from the ASAP7 DRC deck, balanced to one representative case per layer per family where the rule admits the layer. Rule families include minimum horizontal/vertical width, minimum horizontal/vertical spacing, minimum enclosed/overlap area, layer-pair extension, and ASAP7-specific quantization rules (e.g., \texttt{ACTIVE/SDT vertical width must be an integer multiple of 27\,nm}). Two rules from the deck are excluded after pruning topologically equivalent variants; the released \texttt{statistic.json} enumerates the per-rule and per-layer task assignments.
For L1 construction, examples of topology-specific rules that are hand-drawn by physical-design engineers include \texttt{NSELECT.GATE.EX.1-2} and \texttt{PSELECT.GATE.EX.1-2}, which require arranging selector and gate shapes to expose the intended extension violation. Examples of rules that admit parametric templates include \texttt{WELL.S.2} and \texttt{WELL.W.2}, where \texttt{create\_errors.py} instantiates simple rectangles with controlled spacing or width below the ASAP7 threshold.

\subsection{Built-in Innovus DRC repair study}
\label{app:innovus-repair-study}
To verify that DRC-Reasoning does not simply stop before commercial tools finish, we applied Innovus built-in DRC/ECO repair to five representative post-route designs and then re-ran sign-off DRC. As shown in \cref{tab:innovus-repair-study}, built-in repair substantially reduces but does not eliminate violations; DRC-Reasoning is constructed from this residual class of violations that still require manual or agent-driven geometric repair.
\begin{table}[h]
\centering
\caption{\textbf{Residual DRC violations after Innovus built-in repair.}}
\label{tab:innovus-repair-study}
\begin{tabular}{lccccc}
\hline
Design & gcd & aes & ibex & jpeg & riscv \\ \hline
Original & 34 & 972 & 1,551 & 4,481 & 9,395 \\
After Innovus repair & 22 & 760 & 796 & 1,815 & 2,750 \\ \hline
\end{tabular}
\end{table}

\subsection{DRC-Reasoning step-count labelling protocol}
\label{app:step-protocol}
The initial step-count labels that partition DRC-Reasoning into L1 and L2 were produced by a physical-design engineer with 5+ years of industrial sign-off experience and then cross-verified by two physical-design engineers. The reference-labelling procedure is:
\begin{enumerate}[leftmargin=*]
    \item The engineer loads the violation in KLayout and inspects the offending geometry alongside its surrounding layers.
    \item The engineer manually constructs the shortest fix sequence using only the editing operations exposed to the agent (\texttt{add\_shape}, \texttt{change\_shape}, \texttt{move\_cell}); inspection-only tool calls are not counted.
    \item The integer count of editing tool calls is recorded as the case's step count.
    \item After a one-week delay the engineer re-labels the same case without access to the prior label; cases with disagreeing labels are re-evaluated under the protocol until convergence, and any case that still admits competing minimal fixes is retained at the smaller step count.
\end{enumerate}
\paragraph{Two-engineer cross-verification protocol.}
Two physical-design engineers completed cross-verification of the L1/L2 labels. Each independently inspected the cases and constructed a shortest valid edit sequence using the same agent operations, without access to the other engineer's step count. They compared counts and replayed their proposed repairs under the same KLayout sign-off deck. Differences were resolved by checking sequence validity and testing whether a shorter valid sequence existed; the smallest verified edit count was retained. Both initial counts, proposed edit sequences, and the agreed count were recorded with the task metadata.

\subsection{LVS-based audit of functional preservation}
\label{app:drc-functional-audit}
\paragraph{Extraction and comparison.}
To check whether successful DRC repairs preserve circuit connectivity and devices, we wrote a new verification script using \textsc{KLayout} v0.30.3. For each audited task-run, the script uses the two GDS files already retained by the evaluation harness: the golden input \texttt{top\_cell.gds} supplied to the agent and its output \texttt{modified.gds}. It extracts a netlist from each layout with \texttt{pya.LayoutToNetlist} over the ASAP7 conductor/via stack, then performs the LVS netlist-comparison step using the golden layout's extracted netlist as the reference schematic. Each net is keyed by the set of \texttt{(instance, pin)} terminals it connects. The comparison therefore detects opens, shorts, and via deletions that alter connectivity, while tolerating coordinate changes such as wire widening that preserve those terminal connections. The script also checks the device census.

\paragraph{Decision policies.}
We evaluate two policies. The \emph{connectivity/device-evidence policy} declares a break when the extracted netlist or device census changes. The \emph{strictest policy} additionally rejects any movement of an instance classified as a signal cell, regardless of whether its connectivity changes. This extra displacement rule provides a conservative check on geometry edits.

\paragraph{Results and effect on reported scores.}
We audited every archived DRC-successful task-run for Gemini-3-Flash, GPT-5, and Gemma-4-31B-it, totaling $648$ task-runs. The connectivity/device-evidence policy detected $0/648$ breaks ($0.00\%$); the strictest policy flagged $2/648$ runs ($0.31\%$). Both flags correspond to the same edit: a \texttt{VIA\_VIA12} instance moved by $3$\,nm. In both runs, extraction yields $461$ nets before and after the edit, with zero lost nets, zero new nets, and an equal result from \texttt{NetlistComparer}; the flags are caused solely by the additional movement rule. Elsewhere in the corpus, the strictest policy flags $116$ circuit-breaking edits, all in runs that DRC scoring had already rejected.

Thus, no connectivity or device changes were detected among the $648$ audited DRC successes. Recomputing SR with the additional connectivity/device-evidence requirement reproduces the archived success rates to four decimal places, leaving the reported results unchanged for these audited runs.

\section{PPA-Bench: Reproducibility, Tool Contracts, and Construction Details}
\label{app:ppa-reproducibility}

\subsection{Software stack and tool versions}
PPA-Bench is reproduced end-to-end with the following pinned stack:
\begin{itemize}[leftmargin=*]
    \item \textbf{PDK:} ASAP7~\cite{clark2016asap7}, identical to the DRC-Bench setup.
    \item \textbf{Flow:} \textsc{OpenROAD-flow-scripts}, pinned to commit \texttt{a2bb042b6} (release tag \texttt{26Q1-127-ga2bb042b6}). The flow is invoked inside a Docker container so that local installations cannot perturb scoring.
    \item \textbf{Eval driver:} \texttt{eval/eval\_ppa\_sr.sh}. Default configuration: \texttt{TEMPERATURE=0}, \texttt{NUM\_RUNS=5}, \texttt{TOOL\_TIMEOUT=1800\,s} (30-minute wall-clock cap per \texttt{run\_openroad} call), \texttt{MAX\_PARALLEL\_AGENTS=5}. The per-task tool-call iteration limit is varied as an evaluation hyperparameter rather than fixed.
    \item \textbf{Scorer:} \texttt{benchmark/utils/ppa\_score.py}, which implements the SR / NIS computation described in Eq.\,\ref{eq:ppa-si}--\ref{eq:score-ppa}.
    \item \textbf{Agent tools:} released under \texttt{agents/ppa/tools/} (full schemas in \cref{app:ppa-tools}).
\end{itemize}

\subsection{Per-task layout}
\label{app:ppa-manifest}
Each PPA-Bench task ships as a self-contained project directory:
\begin{itemize}[leftmargin=*]
    \item \texttt{src/}: the design's Verilog sources.
    \item \texttt{config/} (a.k.a.\ \texttt{config.mk}, \texttt{constraints.sdc}): OpenROAD configuration and SDC constraints, the primary surface the agent edits.
    \item \texttt{logs/}, \texttt{reports/}, \texttt{objects/}, \texttt{results/}: the previous-iteration intermediate artifacts produced by OpenROAD.
    \item \texttt{prompt.txt}: the natural-language prompt summarizing current PPA values, target thresholds, any constraint floors, and a pointer to \texttt{FlowVariables.md}.
    \item \texttt{info.json}: ground-truth metadata (initial values, targets, reference knob/value records). Read access by the agent is blocked at the \texttt{read\_file} layer so that targets cannot be trivially reached by reading the reference fix; the file is retained on disk for auditability and scoring.
\end{itemize}
Tasks are organized as \texttt{benchmark/L3/asap7/single\_dimension/openroad/\{area, performance, power\}/\{class1,2,3\}} for PPA-Mono and \texttt{benchmark/L3/asap7/multi\_dimension/openroad/\{class1,2\}} for PPA-Multi.

\subsection{Curated knob list for the parameter sweep}
\label{app:ppa-knobs}
The grid sweep used for PPA-Mono and PPA-Multi construction draws from a curated subset of \texttt{eval/FlowVariables.md}. Representative knobs include \texttt{ABC\_AREA}, \texttt{ASAP7\_USE\_VT}, \texttt{CORE\_UTILIZATION}, \texttt{PLACE\_DENSITY}, \texttt{SYNTH\_HIERARCHICAL}, \texttt{TNS\_END\_PERCENT}, \texttt{ROUTE\_EFFORT}, and \texttt{REPAIR\_PDN\_VIA\_LAYER}. The complete list and the per-knob value ranges used in our sweep are released alongside the benchmark.

\subsection{Agent tool contracts}
\label{app:ppa-tools}
The four PPA-Bench tools have the following contracts (full Python signatures and return schemas are released under \texttt{agents/ppa/tools/}).
\begin{itemize}[leftmargin=*]
    \item \texttt{read\_file(command)}: executes a shell-style read command inside a per-task sandbox. A whitelist of safe commands is permitted (\texttt{cat}, \texttt{head}, \texttt{tail}, \texttt{grep}, \texttt{awk}, \texttt{sed}, \texttt{wc}, \texttt{sort}, \texttt{uniq}, \texttt{cut}, \texttt{tr}, \texttt{ls}, \texttt{find}, \texttt{tree}, \texttt{jq}, \texttt{less}, \texttt{more}, \texttt{file}, \texttt{stat}, \texttt{diff}, \texttt{strings}, \texttt{xxd}, \texttt{od}); modifying commands and shell control operators (\texttt{>}, \texttt{>>}, \texttt{\&\&}, command substitution, etc.) are blocked. Pipes are allowed. Reads of \texttt{info.json} or paths outside the sandbox are rejected.
    \item \texttt{edit\_file(path, old\_content, new\_content)}: in-place replacement edit on files inside the sandbox. A small set of structural variables in \texttt{.mk} files (\texttt{PLATFORM}, \texttt{DESIGN\_NAME}, \texttt{DESIGN\_NICKNAME}, \texttt{VERILOG\_FILES}, \texttt{SDC\_FILE}) is locked to prevent flow-breaking modifications; all other configuration entries, SDC constraints, and Verilog sources are editable.
    \item \texttt{run\_openroad(\dots)}: launches an OpenROAD flow run on the current sandbox, returning stdout/stderr and exit status. Each call is bounded by a \texttt{TOOL\_TIMEOUT} environment variable. A flow crash returns the error to the agent and counts as one iteration.
    \item \texttt{report\_ppa()}: parses the latest reports and returns a structured summary of effective period, total power, and die area together with the gaps to the per-task targets.
\end{itemize}

\subsection{Testbench-based functional checks}
\label{app:func-equiv}
PPA-Bench evaluates target attainment together with each task's applicable testbench-based functional checks. The check in \texttt{eval/equiv/check\_equiv.sh} compares the original and modified RTL outputs while accounting for cycle-latency and handshake-timing changes introduced by pipelining, retiming, and datapath restructuring.

For designs with a supplied, simulator-compatible testbench, we compile and simulate the original and modified sources with \textsc{Icarus Verilog} (\texttt{IVERILOG\_LANG=2012}). The testbench at \texttt{eval/equiv/tb/<design>\_tb.v} drives a representative input sequence and emits cycle-stamped output traces for both versions. Traces are compared in one of two modes, selected per design:
\begin{itemize}[leftmargin=*]
    \item \textbf{Event-matched mode.} For combinational and untimed designs, traces are diffed line-by-line; any divergence fails the check.
    \item \textbf{Latency-sweep mode.} For pipelined or handshake-timed designs, a separate comparator (\texttt{eval/equiv/compare.py}) sweeps a latency offset $L \in [0, L_{\max}]$ (default $L_{\max}=16$ cycles, with an $8$-cycle reset warmup) and checks whether $\text{orig}[t] = \text{mod}[t-L]$ over the overlap window for some $L$. The check passes if any offset yields zero mismatches, so legitimate pipeline-depth or handshake-timing changes do not count as functional regressions.
\end{itemize}
Task-level records identify testbench applicability and retain the trace-comparison outcome or build/error status.

\section{Agent Frameworks and Experimental Protocol}
\label{app:agent-frameworks}

\subsection{Models and inference setup}
\label{app:models}
The nine models reported in \cref{tab:overall-performance} are evaluated under a shared agent codepath, with the model identifiers, inference settings, and tool-call accounting documented below.
\begin{itemize}[leftmargin=*]
    \item \textbf{Commercial.} GPT-5, GPT-5-mini, Gemini-3-Flash-preview, and Claude Opus 5, accessed via their official APIs at the time of evaluation. The exact API model identifiers are recorded in the per-run cost-tracker logs released with the benchmark. Claude uses the identifier \texttt{claude-opus-5}, with the temperature parameter omitted from API requests.
    \item \textbf{Open-source.} DeepSeek-V3.2, Qwen3.5-122B-A10B, Qwen3.5-27B, Gemma-4-31B-it, and Qwen3.5-9B, each served via \textsc{vLLM} on a local GPU cluster behind an OpenAI-compatible \texttt{/v1} endpoint. Tensor-parallel degree is selected per checkpoint to fit GPU memory; greedy decoding (\texttt{temperature=0}) is enforced server-side.
\end{itemize}
Token usage and tool-call counts are tracked per run via a shared cost-tracking utility shipped alongside the agent code.

\paragraph{Repeated runs and inference reproducibility.}
Each evaluated configuration is run five times from the same task inputs, using temperature zero where supported and the Claude API settings specified above. We report run-averaged scores and retain per-run outcomes and tool traces; uncertainty is assessed at the task level in \cref{app:variance}.

\paragraph{Runtime and API-cost accounting.}
\label{app:runtime-cost}
The runtime and cost panels in \cref{tab:runtime-cost} summarize the archived evaluations. Elapsed times were measured on one $128$-core node with five agents running concurrently: a single pass over DRC-Bench takes $1.2$--$1.6$ hours and a pass over PPA-Bench takes $4.4$--$5.0$ hours, with approximately $6.7$ and $23$ hours respectively for the full five-run protocol. Provider API charges are summed from per-task \texttt{cost.json} records for one pass. Charges and totals are reproduced as reported, so rounded components may not sum exactly to the displayed total; the five-run protocol incurs approximately five times the per-pass API charge.

\begin{table}[htbp]
\centering
\small
\caption{\textbf{API cost and elapsed runtime on \textsc{PostEDA-Bench}.} The panels report provider API charges and elapsed benchmark runtime from the archived evaluations.}
\label{tab:runtime-cost}
\textbf{API cost per pass (USD; text-only ReAct)}\par
\begin{tabular}{@{}lrrr@{}}
\toprule
Backbone & DRC (70 tasks) & PPA (75 tasks) & Both (145 tasks) \\
\midrule
GPT-5-mini & 1.64 & 5.89 & 7.54 \\
GPT-5 & 13.33 & 37.42 & 50.75 \\
Claude Opus 5 & 50.74 & 184.38 & 235.12 \\
Gemma-4 / Qwen3.5 (self-hosted) & 0 & 0 & 0 \\
\bottomrule
\end{tabular}
\par\smallskip
\textbf{Elapsed runtime: one 128-core node, five concurrent agents}\par
\begin{tabular}{@{}lrr@{}}
\toprule
Suite & One pass (hours) & Five-run protocol (hours) \\
\midrule
DRC-Bench (70 tasks) & 1.2--1.6 & $\sim 6.7$ \\
PPA-Bench (75 tasks) & 4.4--5.0 & $\sim 23$ \\
\bottomrule
\end{tabular}
\par\smallskip
{\footnotesize\raggedright
API charges are archived per-pass measurements for the listed ReAct backbones,
separate from local CPU/GPU expenses. The five-run protocol costs approximately
$5\times$ the listed API charges. Charges and totals are reproduced as reported; rounded components may not sum exactly to the total.
Elapsed times correspond to benchmark passes on the hardware specified above.
See \cref{app:runtime-cost,app:claude-opus,app:versioning} for accounting,
Claude's tool-use profile, and release versioning.\par}
\end{table}

Verification dominates wall-clock time. A \texttt{run\_openroad\_flow} call has a median duration of $271$--$315$ seconds across the measured configurations (90th percentile: $1{,}807$ seconds), compared with $2$--$8$ seconds for an LLM turn, and accounts for $83$--$93\%$ of PPA episode time. A DRC check has a median duration of $62$--$69$ seconds, reaching $481$ seconds on the largest designs, compared with $5$--$10$ seconds for a geometry turn. KLayout accounts for $37$--$61\%$ of DRC wall-clock time for Gemini-3-Flash and Gemma-4. The corresponding OpenROAD call costs roughly as much time as $35$--$158$ LLM turns. These measurements explain why cost and runtime should be reported alongside success rates, and why the relative value of reasoning and additional tool feedback depends on the task family.

\paragraph{Claude Opus 5 results and tool-use profile.}
\label{app:claude-opus}
Claude has the highest observed DRC scores in \cref{tab:overall-performance}: Essential SR/VRR is $95.00/95.17$ and Reasoning SR/VRR is $54.67/54.67$. The gap remains on individual levels: Essential L3 SR is $98.00\%$, whereas Reasoning L2 SR is $36.00\%$. Its PPA-Mono SR/NIS is $40.57/47.69$, and PPA-Multi is $2.00/-1.39$; Performance/Area/Power SR is $0/74/68\%$. These measurements use text-only ReAct with the shared action caps. On PPA, Claude averages $14.0$ \texttt{read\_file} calls and $1.33$ flow runs per episode, versus $4.88$ and $2.05$ for Gemma-4. Each dispatched tool call in a parallel batch counts toward the $16$-action cap. These traces characterize how each model allocates its actions between inspection and flow-based verification. No evaluated configuration leads all four task families.

\subsection{Iteration semantics per framework}
\label{app:iteration-semantics}
We adopt a per-framework definition of one iteration so that the iteration cap (8 for DRC-Bench, 16 for PPA-Bench) is interpreted consistently:
\begin{itemize}[leftmargin=*]
    \item \textbf{ReAct.} One iteration is one tool call dispatched by the agent. Pure reasoning turns that do not emit a tool call are not counted.
    \item \textbf{Proposer--Critic.} The proposer drafts a tool call as a textual description; the critic validates or refines the proposal; only the agreed action is dispatched. The full proposer--critic--dispatch triple counts as one iteration; proposal rounds that do not lead to a dispatch are not counted.
    \item \textbf{ORFS-Agent.} ORFS-Agent uses a fixed Bayesian-optimization budget of $6$ outer iterations $\times\,3$ parallel candidates per outer iteration $=18$ OpenROAD candidate runs per task, hardcoded in the wrapper rather than driven by the global iteration cap. Each candidate run is one OpenROAD flow execution; the single LLM call consumed during the search-space-discovery setup phase is not counted against the budget.
\end{itemize}

\paragraph{Adaptive iteration cap.} The headline caps in \cref{sec:exp-setup} ($8$ DRC, $16$ PPA) are stated in units of \emph{agent actions} (tool dispatches), not raw LLM calls. Multi-agent scaffolds that consume additional internal LLM calls per dispatch---e.g., Proposer--Critic's proposer$\to$critic$\to$dispatch triple, or ToT's per-node candidate sampling and judge scoring---therefore receive a proportionally larger total LLM-call budget than ReAct's raw tool-call budget. We scale the underlying LLM-call budget linearly with the number of cooperating agents per dispatch, so that adding more agents to a workflow grants correspondingly more reasoning calls. This adaptive scaling keeps cross-scaffold comparisons about \emph{scaffold value} rather than penalising frameworks for needing more thought per action.

\subsection{Run termination and timeout handling}
\label{app:exp-protocol}
A run terminates when (i) the success criterion is met, (ii) the iteration cap is reached, (iii) the underlying tool times out, or (iv) the LangGraph recursion limit is hit; in all non-success cases the partial-progress metric is computed from the last valid state. The default \texttt{TOOL\_TIMEOUT} is $1800$\,s for OpenROAD flow runs and the equivalent for KLayout DRC. Tool timeouts and flow crashes return the error to the agent and count as one iteration.

\paragraph{Excluded baselines.} Most prior EDA-LLM benchmarks (\cref{tab:benchmark-comparison}) target script generation from natural language rather than PPA convergence or DRC fixing, and their agents do not expose a comparable success criterion; we therefore exclude them from \cref{tab:overall-performance}. ORFS-Agent~\cite{ghose2025orfsagenttoolusingagentschip} is the closest published baseline for PPA optimization and is included via the wrapper described below.

\subsection{ORFS-Agent search-space-discovery wrapper}
\label{app:orfs-wrapper}
The published ORFS-Agent~\cite{ghose2025orfsagenttoolusingagentschip} requires every (design, PDK) pair to ship a hand-curated \texttt{opt\_config.json} that lists the tunable parameters and their ranges, plus per-design baseline ECP and wirelength values. Our PPA-Bench tasks are driven by natural-language prompts and do not carry this metadata, so we adapt the published agent with an LLM-based setup phase that automates the configuration step. Concretely:
\begin{itemize}[leftmargin=*]
    \item A single LLM call reads the task's \texttt{prompt.txt}, the contents of \texttt{config/} (i.e., \texttt{config.mk} and \texttt{constraints.sdc}), the list of files under \texttt{src/}, and \texttt{eval/FlowVariables.md}, and returns a structured search space (tunable parameters, ranges, defaults) via a single tool call. Targets and the optimization objective are not picked by the LLM---they are parsed from \texttt{info.json} by the orchestrator outside the agent's view---so the discovery step affects only \emph{which knobs to sweep}, not the success criterion.
    \item The discovered search space is fed into the original GP-driven outer loop: iteration~1 uses Latin Hypercube initialization around the discovered defaults; iterations~2--$N$ propose new candidates by maximizing a UCB acquisition over a Mat\'ern-kernel GP, with diversity-aware top-$k$ selection. Each candidate runs a real OpenROAD flow inside an isolated per-candidate sandbox.
    \item The evaluation allocates $6$ outer iterations with $3$ parallel candidates per iteration, totaling $18$ OpenROAD candidate runs per task (\cref{app:iteration-semantics}). Each candidate flow execution counts as one ORFS-Agent iteration.
\end{itemize}
The wrapper preserves the upstream agent's CLI contract so that \texttt{eval\_ppa\_sr.sh} can swap it in for ReAct or Proposer--Critic without modification.

\subsection{Framework-level system prompts}
\label{app:agent-prompts}
We release the system prompts used by ReAct and Proposer--Critic for both DRC-Bench and PPA-Bench under \texttt{agents/\{drc,ppa\}/\{react,proposer\_critic\}/}. The prompts are held fixed across models within a framework so that benchmark scores reflect model and scaffold capability rather than prompt engineering. Each prompt declares (a) the role and tool repertoire available to the model, (b) the success criterion (DRC-clean for DRC-Bench, target metrics + applicable testbench-based functional checks for PPA-Bench), and (c) framework-specific scaffolding---chain-of-thought interleaving for ReAct, role separation between proposer and critic for Proposer--Critic, and the structured search-space schema for the ORFS-Agent setup phase.

\subsection{Reflexion and Tree-of-Thought wrappers (PPA)}
\label{app:reflexion-tot}
We also release PPA-side Reflexion and Tree-of-Thought wrappers used in the iteration-budget and thinking-mode ablations (\cref{sec:iter-ablation,sec:thinking-ablation}); both are layered on top of the same ReAct PPA primitives and tool surface so that scores reflect the search/reflection policy rather than tool plumbing.

\paragraph{Reflexion (\texttt{agents/ppa/reflexion/}).} The wrapper implements the verbal-reinforcement trial-and-reflect loop of \citet{shinn2023reflexion}. The user-visible iteration budget is divided evenly across $N$ trials (we use $N{=}2$, matching the halve-and-reflect-once protocol of \cref{sec:iter-ablation}). Each trial is a full ReAct rollout with a freshly initialised conversational memory; between trials the agent makes a tool-less Self-Reflection LLM call that summarises lessons from the previous trajectory, and these lessons are appended to the system prompt of the next trial as long-term memory. The Self-Reflection call does not consume iteration budget. To match canonical Reflexion semantics, the working sandbox is snapshotted before trial 1 and restored before every subsequent trial, so each trial sees the original task content; only the verbal reflection is carried across. Cumulative tool-call counts and token costs are aggregated across trials so the eval driver (\texttt{eval\_ppa\_sr.sh}) sees a single agent-run summary, matching the ReAct-agent contract.

\paragraph{Tree-of-Thought (\texttt{agents/ppa/tot/}).} The wrapper implements the ToT search of \citet{yao2023tot} over the same tool surface as the ReAct PPA agent. At each tree node the agent samples \texttt{PARALLEL\_NODE} candidate next actions in parallel, an LLM judge scores them, the highest-scored candidate is executed, and the resulting state becomes a new node. Un-executed candidates from every previously-expanded node remain in a global priority queue, so when the current path stalls the search backtracks to the highest-scored alternative from an ancestor. Iteration budget is the total tool-call budget across the whole tree; \texttt{max\_depth} caps the deepest single path (default $16$, matching ReAct's per-run cap), and per-step LLM calls (candidate sampling and scoring) are billed in tokens but do not consume iteration budget. Each node owns a sandbox snapshot under \texttt{\$AGENT\_RESULT\_DIR/\_tot\_snapshots/}; on backtrack, the sandbox is wiped and restored from the chosen ancestor's snapshot before executing. At end of run the sandbox is restored to the best node, ranked by (i) success, (ii) at least one successful \texttt{run\_openroad\_flow} call, (iii) deeper paths over shallower ones, and (iv) judge score, so the eval driver picks up a node whose OpenROAD reports actually exist.

\section{Extended Analysis of Main Results}
\label{app:extended-results}
This section expands on the deductions in \cref{sec:main-results}, focusing on framework-level effects and cross-dimension comparisons that are too detailed to fit inline.

\subsection{Scaffold value is dimension-conditioned and reverses sign on PPA}
\label{app:ext-scaffold}
Switching from ReAct to Proposer--Critic raises Qwen3.5-122B's DRC-Essential SR by $+15$ points and Gemma-4-31B-it's DRC-Reasoning SR by $+3.3$, but cuts Gemma-4's PPA-Mono SR by $22$ points ($64.56\!\to\!42.28$) and Qwen3.5-9B's by $21.7$. Since our evaluation uses adaptive, approximately matched budgets across frameworks, this reversal is not explained by fewer allowed optimization steps. Instead, it suggests that proposer--critic validation changes the search policy: on DRC tasks where each call is a local geometric edit, the validator catches malformed edits before dispatch and pays for itself; on PPA tasks, where progress depends on empirical \texttt{run\_openroad} measurements, critic-side filtering can bias the trajectory toward conservative local changes and reduce broad parameter exploration. Framework choice is therefore not a global hyperparameter, and cross-framework comparisons that aggregate over dimensions will mislead.

\subsection{Bayesian optimization amplifies the LLM's search-space prior}
\label{app:ext-orfs}
ORFS-Agent lifts Qwen3.5-122B's PPA-Multi NIS from $35.46$ to $67.80$ (the table's highest value, $\sim 1.9\times$) but drives Gemma-4-31B-it's PPA-Multi NIS from $+16.34$ to $-15.90$ and its PPA-Mono NIS from $69.34$ to $47.31$. Both runs share the same outer BO loop and budget; the only difference is the LLM-induced search space inferred from \texttt{FlowVariables.md}. The Pareto-frontier failure of pure ReAct is therefore best understood as a poor exploration policy layered on largely correct knob knowledge: when a structured sampler is grafted on top of a competent space, multi-objective SR jumps to first place; when it is grafted on top of a wrong space, it amplifies the error rather than correcting it. The open research problem is LLM-side search-space discovery, not the optimization loop itself.

\subsection{Capability-gap ordering on \sysname}
\label{app:ext-ordering}
Top-cell ratios across dimensions give a quantitative difficulty ordering: PPA-Multi reach is $20.00/64.56=31\%$ of PPA-Mono reach, while DRC-Reasoning reach is $54.67/95.00\approx58\%$ of DRC-Essential reach. Multi-objective trade-off is therefore the largest absolute capability gap today, ahead of geometric DRC reasoning. This suggests that the next-step research direction is not better DRC-rule prompting but tool-augmented planners with explicit constraint-tracking memory.

\subsection{Additional PPA-Bench per-level breakdowns}
\label{app:ppa-perlevel}
\begin{figure*}[t]
  \centering
  \begin{subfigure}[b]{0.32\textwidth}
    \centering
    \includegraphics[width=\linewidth]{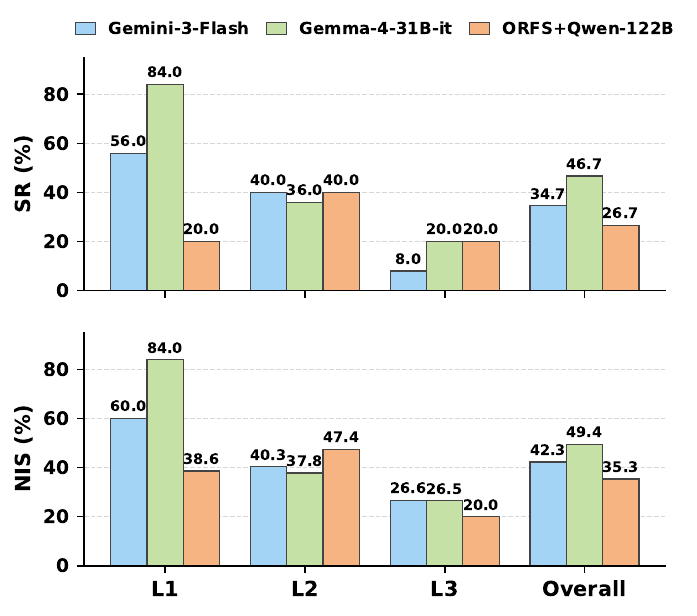}
    \caption{PPA-Performance (L1/L2/L3).}
    \label{fig:ppa-perf-full}
  \end{subfigure}
  \hfill
  \begin{subfigure}[b]{0.32\textwidth}
    \centering
    \includegraphics[width=\linewidth]{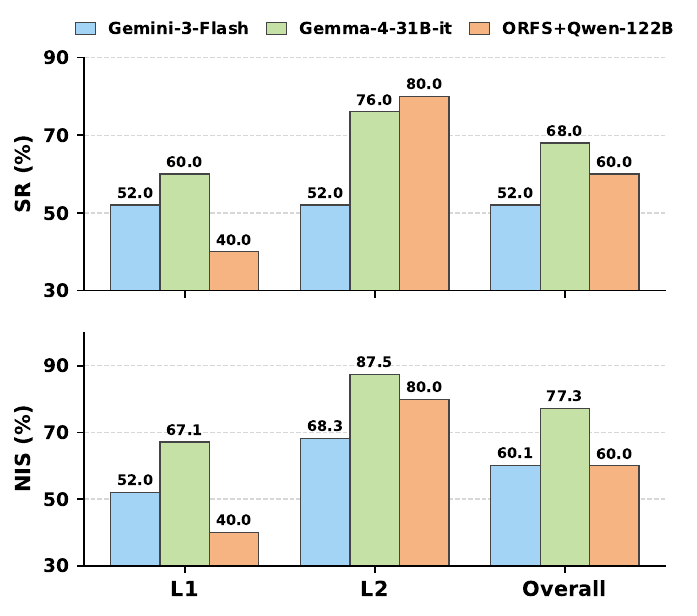}
    \caption{PPA-Power (L1/L2).}
    \label{fig:ppa-power-full}
  \end{subfigure}
  \hfill
  \begin{subfigure}[b]{0.32\textwidth}
    \centering
    \includegraphics[width=\linewidth]{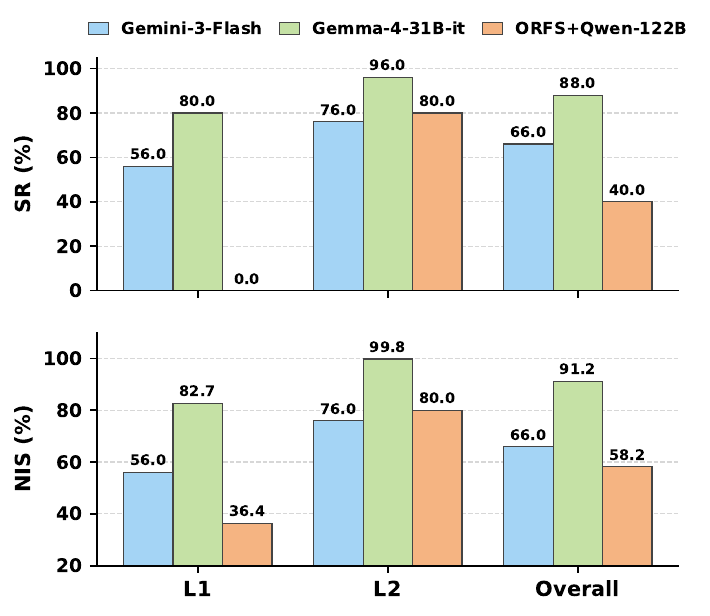}
    \caption{PPA-Area (L1/L2).}
    \label{fig:ppa-area-full}
  \end{subfigure}
  \\[6pt]
  \begin{subfigure}[b]{0.49\textwidth}
    \centering
    \includegraphics[width=\linewidth]{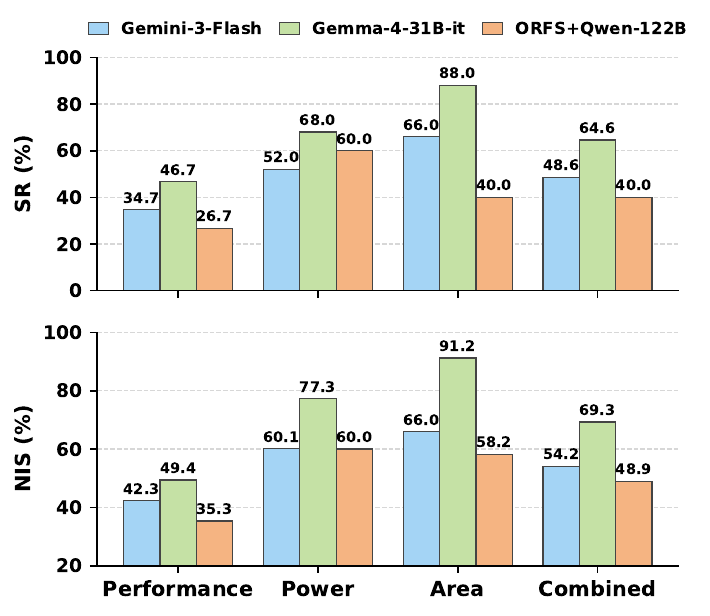}
    \caption{PPA-Mono (per sub-dimension + Combined).}
    \label{fig:ppa-mono-full}
  \end{subfigure}
  \hfill
  \begin{subfigure}[b]{0.49\textwidth}
    \centering
    \includegraphics[width=\linewidth]{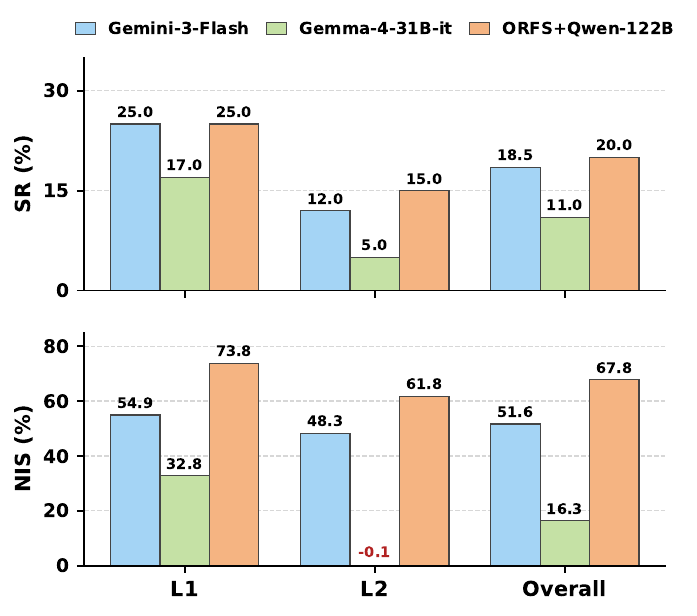}
    \caption{PPA-Multi (L1/L2).}
    \label{fig:ppa-multi-full}
  \end{subfigure}
  \caption{\textbf{Full PPA-Bench per-level performance.} SR (top) and NIS (bottom) per construction-time level. Row~1 decomposes PPA-Mono into Performance / Power / Area; row~2 summarizes PPA-Mono per sub-dimension and reports PPA-Multi.}
  \label{fig:ppa-perlevel-full}
\end{figure*}

\Cref{fig:ppa-perlevel-full} gives the full five-panel PPA decomposition used to support the compact main-text table in \cref{tab:ppa-bench}.

\textbf{Power/Area show inverted L1$\to$L2.} Both PPA-Power and PPA-Area have $\mathrm{L2}\!\ge\!\mathrm{L1}$ SR for every model (e.g., ORFS+Qwen $40\!\to\!80\%$ on Power and $0\!\to\!80\%$ on Area; Gemma $60\!\to\!76\%$ on Power). This counter-intuitive ordering suggests that joint power/area perturbations create slack: additional misconfigured knobs offer more recovery levers, whereas single-knob cases require identifying the responsible knob exactly.

\textbf{The full decomposition clarifies ORFS behavior.} ORFS+Qwen leads PPA-Multi NIS by a wide margin ($67.80$ vs Gemini $51.62$ and Gemma $16.34$) and matches or exceeds the others on Power L2 / Area L2, but trails the PPA-Mono Combined score ($40.00/48.91$ vs Gemma $64.56/69.34$) and has $0\%$ SR on Area L1. This supports the main-text conclusion that Bayesian optimization helps through exploration coverage when the inferred search space contains useful knobs, but can amplify a weak inferred space when the task hinges on a specific single knob.

\section{Detailed Ablation Readings}
\label{app:ablation-tables}
We expand on the ablation figures in \cref{sec:vision-ablation,sec:iter-ablation,sec:thinking-ablation}. Source tables and per-run logs accompany the released benchmark.

\paragraph{Detailed iteration-budget reading.}
\Cref{fig:iter-ablation} separates two regimes. On DRC-Essential and PPA-Mono, SR follows a concave shape ($59.5\!\to\!82.0\!\to\!95.0$ and $25.1\!\to\!64.6\!\to\!74.9$), so extra iterations mostly polish already reachable tasks. DRC-Reasoning and PPA-Multi behave oppositely: Reasoning gains only $+5.3$ SR from $4\!\to\!8$ but $+15.3$ from $8\!\to\!16$, while Multi gains $+10$ from $8\!\to\!16$ and $+17$ from $16\!\to\!32$. The PPA-Multi NIS$-$SR gap also narrows then widens ($10.91\!\to\!5.34\!\to\!12.04$), indicating that additional iterations can improve the violated metric without crossing the success boundary. DRC-Essential shows the opposite pattern: its VRR$-$SR gap collapses from $9.75\!\to\!10.0\!\to\!0.42$, consistent with all-or-nothing cascaded fixes once enough edit budget is available.

\section{Variance and Statistical Uncertainty}
\label{app:variance}

\paragraph{Task-level paired bootstrap.}
We quantify uncertainty with a paired cluster bootstrap over tasks using
$20{,}000$ resamples. For each task, success is averaged over its five runs.
Tasks are resampled as clusters, retaining their repeated-run outcomes and
using the same resampled task indices for both configurations in a comparison.
This preserves pairing by task rather than treating all task-runs as independent
observations. We report differences in SR in percentage points and their
$95\%$ confidence intervals. The bootstrap uses the underlying task-level
outcomes rather than the rounded means displayed in the main table.

\begin{table}[htbp]
\centering
\small
\caption{\textbf{Paired SR differences and uncertainty.} Gemini denotes
Gemini-3-Flash-preview, Gemma-4 denotes Gemma-4-31B-it, and Qwen-122B denotes
Qwen3.5-122B-A10B. Models use ReAct unless ORFS-Agent is explicitly indicated.
Differences are computed as the first configuration minus the second.}
\label{tab:paired-variance}
\begin{tabular}{@{}lrr@{}}
\toprule
Paired comparison & $\Delta$ SR (pp) & $95\%$ CI (pp) \\
\midrule
DRC-Essential: Gemini $-$ GPT-5 & $+1.00$ & $[-4.00, +6.50]$ \\
DRC-Reasoning: Gemini $-$ GPT-5 & $+5.33$ & $[-5.33, +16.67]$ \\
DRC-Essential: Gemini $-$ Qwen-122B & $+34.00$ & $[+23.00, +45.50]$ \\
PPA-Mono: Gemma-4 $-$ GPT-5 & $+25.14$ & $[+14.29, +37.14]$ \\
PPA-Multi: ORFS+Qwen $-$ Qwen-122B & $+16.00$ & $[+3.50, +29.50]$ \\
\bottomrule
\end{tabular}
\end{table}

The near-ties in the first two rows of \cref{tab:paired-variance} have intervals
containing zero and do not support a statistical ordering between the paired
models. The larger separations in the remaining rows exclude zero. Thus the
bold and underlined entries in \cref{tab:overall-performance} describe observed
means, while the absence of a single leading configuration is a cross-dimension
observation. These bootstrap comparisons concern the archived baseline and
vision configurations.

\paragraph{Uncertainty of the vision effect.}
Of the eight paired SR improvements from adding vision to four backbones on
DRC-Essential and DRC-Reasoning, only two individual intervals exclude zero:
GPT-5 on Essential and Qwen3.5-122B-A10B on Essential
(\cref{tab:vision-uncertainty}). When pooled over the four backbones, the SR
improvement is positive on both task families. The probability that all eight
SR differences are non-negative under this analysis is $0.31$. Consequently,
the non-decreasing SR/VRR values across all $16$ observed metric cells are a
description of these runs, not evidence of a universal or statistically
significant improvement for every backbone and metric.

\begin{table}[htbp]
\centering
\small
\caption{\textbf{Vision-minus-text SR differences.} The first two rows are the
individual model--family comparisons whose $95\%$ intervals exclude zero;
the remaining rows pool the four evaluated backbones.}
\label{tab:vision-uncertainty}
\begin{tabular}{@{}lrr@{}}
\toprule
Comparison & $\Delta$ SR (pp) & $95\%$ CI (pp) \\
\midrule
GPT-5, DRC-Essential & $+9.50$ & $[+2.00, +18.50]$ \\
Qwen3.5-122B-A10B, DRC-Essential & $+13.50$ & $[+6.50, +20.50]$ \\
Pooled, DRC-Essential & $+6.37$ & $[+3.38, +9.62]$ \\
Pooled, DRC-Reasoning & $+5.50$ & $[+2.33, +9.00]$ \\
\bottomrule
\end{tabular}
\end{table}

\paragraph{Synthetic-to-practical gap.}
For Gemini-3-Flash-preview, the DRC-Essential minus DRC-Reasoning SR gap is
$48.83$ percentage points, with a $95\%$ interval of $[29.50, 67.17]$.
This supports the cross-family difficulty gap despite uncertainty in nearby
within-column model rankings. Per-run, per-task outcomes for all $16$ evaluated
DRC configurations and the corresponding PPA records will be released to
support independent aggregation and uncertainty analysis.


\end{document}